\documentclass[twocolumn,secnumarabic, nobibnotes, aps, prb]{revtex4-2}

\usepackage[english]{babel}

\usepackage{fullpage}

\usepackage[utf8]{inputenc} % enables inputs like é and ö
\usepackage{lmodern} % better font rendering
\usepackage[T1]{fontenc}

\usepackage{amssymb}
\usepackage{mathtools} % extension of amsmath
	\numberwithin{equation}{section}

\usepackage[bb=dsserif,bbscaled=1.15]{mathalpha}
\usepackage{verbatim} % e.g. \begin{comment}

\usepackage[dvipsnames]{xcolor} % for colours used for hyperref below

\usepackage{graphicx}
\usepackage{graphics}
\usepackage{pgfplots}

\usepackage{tikz}
	\usetikzlibrary{decorations.pathreplacing}
	\usetikzlibrary{decorations.pathmorphing} % zigzag lines
	\usetikzlibrary{matrix,arrows} % commutative diagrams in tikz
	\usetikzlibrary{patterns} % hatching
	\usetikzlibrary{arrows.meta} 

\usepackage{multirow} % \multirow, \multicol

\usepackage{enumitem} % e.g. [label=\emph{\roman*}), resume]
	\setlist[itemize]{left= \parindent .. 2\parindent, label=\raisebox{0.14ex}{\scriptsize$\bullet$}} % cf tex.stackexchange.com/a/119322/

\usepackage{scalerel,stackengine} % between big an Big https://tex.stackexchange.com/a/483979/

\makeatletter
	\newcommand\niton{\mathrel{\m@th\mathpalette\canc@l\owns}}
	\newcommand\canc@l[2]{{\ooalign{$\hfil#1/\mkern1mu\hfil$\crcr$#1#2$}}}
	\makeatother

\DeclarePairedDelimiter{\ket}{\lvert}{\rangle}
\DeclarePairedDelimiterX{\ketbra}[2]{\lvert}{\rvert}{#1\rangle \langle#2}
\DeclarePairedDelimiterX{\braket}[2]{\langle}{\rangle}{#1\vert#2}

\DeclarePairedDelimiterX{\cbraket}[2]{\langle\!\langle}{\rangle}{#1\vert#2}
\DeclarePairedDelimiterX{\bracket}[2]{\langle}{\rangle\!\rangle}{#1\vert#2}
\DeclarePairedDelimiterX{\cbracket}[2]{\langle\!\langle}{\rangle\!\rangle}{#1\vert#2}

\usepackage{bm} % cf https://tex.stackexchange.com/a/10643
	\newcommand{\vect}[1]{\bm{{#1}}}

\usepackage{braket, relsize, nccmath} % cf https://tex.stackexchange.com/a/166548/

\newcommand{\I}{\mathrm{i}}
\newcommand{\E}{\mathrm{e}}

\newcommand{\bd}{\begin{displaymath}}
	\newcommand{\ed}{\end{displaymath}}
\newcommand{\be}{\begin{equation}}
	\newcommand{\ee}{\end{equation}}
\newcommand{\bea}{\begin{eqnarray}}
	\newcommand{\eea}{\end{eqnarray}}
\newcommand{\C}{\mathbb{C}}
\newcommand{\R}{\mathbb{R}}

\newcommand{\Z}{\mathbb{Z}}
\newcommand{\e}{\epsilon}

 \let\Im\undefined \DeclareMathOperator{\Im}{Im}

\DeclareMathOperator*{\ordprod}{\prod\limits^{\vbox to -.5ex{\kern-0.5ex\hbox{$\leftharpoonup$}\vss}}}
\DeclareMathOperator*{\ordprodopp}{\prod\limits^{\vbox to -.5ex{\kern-0.5ex\hbox{$\rightharpoonup$}\vss}}}

\makeatletter
\DeclareRobustCommand\widecheck[1]{{\mathpalette\@widecheck{#1}}}
\def\@widecheck#1#2{%
	\setbox\z@\hbox{\m@th$#1#2$}%
	\setbox\tw@\hbox{\m@th$#1%
		\widehat{%
			\vrule\@width\z@\@height\ht\z@
			\vrule\@height\z@\@width\wd\z@}$}%
	\dp\tw@-\ht\z@
	\@tempdima\ht\z@ \advance\@tempdima2\ht\tw@ \divide\@tempdima\thr@@
	\setbox\tw@\hbox{%
		\raise\@tempdima\hbox{\scalebox{1}[-1]{\lower\@tempdima\box
				\tw@}}}%
	{\ooalign{\box\tw@ \cr \box\z@}}}
\makeatother
\usepackage{marginnote}

\usepackage{xurl} %https://tex.stackexchange.com/questions/3033/forcing-linebreaks-in-url

\usepackage[
]{hyperref} % after most other packages; ocgcolorlinks cf tex.stackexchange.com/questions/4425/

\usepackage[ocgcolorlinks]{ocgx2} %	https://github.com/latex3/hyperref/issues/31

\hypersetup{linkcolor=BlueViolet, citecolor=OliveGreen, urlcolor=RawSienna, %filecolor=Sepia
}

\newcommand{\gz}{{\color{lightgray}{0}}}

\begin{document}
	
	\title{Fukui--Kawakami chains: spectrum and hidden $\mathfrak{gl}(1|1)$-symmetry}
	
	\author{Rob Klabbers$^a$ and Antoine Lefebvre$^b$ }
	\address{%
		{\vspace{0.1cm}$^{a}$\,Humboldt-Universität zu Berlin,~Zum Großen Windkanal 2, 12489 Berlin, Germany}\\
		{\vspace{0.1cm} $^{b}$\,Laboratoire de Physique de l'ENS, Sorbonne Université,~24 rue Lhomond, 75005 Paris, France}
		}
	
\date{\today}

\begin{abstract} 
 Fukui and Kawakami showed that the trigonometric Haldane--Shastry (HS) spin chain can be deformed by introducing twisted boundary conditions. In this work we revisit the resulting twisted HS chains, which we call Fukui--Kawakami (FK) chains. 
 
 We analyse their spectra in dependence of the twist parameter, utilising a direct connection with the (untwisted) HS chain. This allows us to explain why part of the spectrum can be described by Fukui and Kawakami's twisted Bethe equations, and can be constructed from Yangian highest weight states. These states can be labelled by combinatorial data called `motifs', as in the HS chain, which cover some of the (deformed) descendants. We furthermore show that there are other descendants which do not follow from these Bethe equations, but whose energy can be described as a sum of two single-particle energies, suggesting additional hidden symmetries. 
 
 We then focus on the special case of antiperiodic boundary conditions, and show that this chain coincides with the `minimally polarised' long-range model recently introduced by Basu-Mallick, Finkel, and González-López. Remarkably, this connection implies that the antiperiodic chain has a hidden $\mathfrak{gl}(1|1)$-symmetry, which we use to relate the HS motifs to their `supersymmetric' $\mathfrak{gl}(1|1)$ counterparts.  
\end{abstract}

\hfill HU-EP-26/28-RTG
	
\maketitle
\tableofcontents

\section{Introduction}
Integrable spin chains are great theoretical laboratories to study physical phenomena in magnetic materials, combining simplicity with powerful analytic methods. The simplest models of this kind are short-range integrable chains such as the Heisenberg \textsc{xxx} model, for which the toolkit is most developed, see e.g. \cite{Maillet2007}. 

An important technique is \emph{twisting} \footnote{In this work we only consider diagonal (or abelian) twisting}. For nearest-neighbour chains twisting can be thought of as adding a defect in between the first and last sites of a periodic chain to deform the boundary conditions, introducing a free parameter such that the model can be applied to a wider variety of realistic scenarios. On a technical level, twisting also resolves degeneracies in the spectrum \cite{Maillet2007}, which makes the (algebraic) Bethe ansatz more effective and paves the way for proofs of its validity; that each solution of the Bethe ansatz equations describes a physical state (faithfulness) and that each physical state is described by such a solution (completeness), see \cite{chernyakCompletenessWronskianBethe2022} and references therein. 

Experimental cold-atom setups, such as Rydberg atom arrays or atomic ion traps, often exhibit long-range interactions, and their properties are naturally modelled by chains with such interactions. Indeed, integrable long-range spin chains such as the Haldane--Shastry (HS) chain have been fruitful models in the study of e.g. fractional statistics \cite{Hal_91b} and ballistic transport \cite{sirkerConservationLawsIntegrability2011,bulchandaniHydrodynamicsSpinTransport2024}. 

Here we aim to further develop an understanding of twisting for such long-range chains. In the absence of a full algebraic framework -- such as the Sklyanin's boundary Yang-Baxter equation \cite{sklyaninBoundaryConditionsIntegrable1988} or Drinfel'd twists \cite{drinfeldQuasiHopfAlgebrasKnizhnikZamolodchikov1989} for short-range integrability -- a direct way to obtain a twisted chain is by wrapping \cite{sutherlandExactResultsQuantum1971}: putting an infinitely-long system on a finite volume by periodising its pair interaction under twisted boundary conditions \cite{klabbers2024landscapes}. Using this method Fukui and Kawakami \cite{fukui1996exact} constructed and subsequently studied a family of spin chains that deform the HS chain and we will call the Fukui--Kawakami (FK) chains. These twisted chains still retain some of the extended (Yangian) symmetry which drives the integrability of the HS chain, and their spectra are (at least partially) described by a Jastrow ground-state wavefunction and twisted version of Haldane's Bethe-ansatz-like equations \cite{fukui1996exact}. Nevertheless, the question of integrability of the FK chains remains open; there is no known set of commuting hamiltonians and no construction of further symmetries, e.g. cf. \cite{bernard1993yang}.

In this work, we study the FK chains in more detail: after reviewing their construction by twisted wrapping, we use a trigonometric identity to relate them directly to the HS chain. Using this relation we construct part, but not all, of the spectrum, and explain why Haldane's Bethe-like equations show up and can be used to label states using the Yangian motifs of the HS chain. We make progress on an alternative description for the missing energies by a study of the two-magnon sector, and show in particular that the energy of these descendants can be interpreted as a sum of two single-particle energies. We then turn to the special case of antiperiodic twisting, and discuss how this model has appeared in the literature in several forms, including through an \textit{R}-matrix-valued Lax pair construction \cite{sechin2018r}. We prove that it is equivalent to the minimally polarised long-range model recently proposed in \cite{basu-mallickNovelClassTranslationally2020}. Since the latter is also equivalent to the supersymmetric HS chain, this proves that the antiperiodic chain has a hidden $\mathfrak{gl}(1|1)$-symmetry. We further highlight this by showing how this chain of equivalences directly relates the HS-chain's ($\mathfrak{gl}(2|0)$-)motifs to their supersymmetric ($\mathfrak{gl}(1|1)$-)motif counterparts through a `thickening' procedure. We delegate a proof of the central trigonometric identity and a description of the supersymmetric HS chain to appendices. 
\newpage
\section{Wrapping}
\label{sec:twisted_models}
Wrapping can be used to construct both periodic and twisted chains from those defined on an infinite one-dimensional lattice. We review how in the periodic case this results in the usual periodic Heisenberg \textsc{xxx} chain and its long-range counterpart -- the Haldane--Shastry chain -- and we recall its pertinent properties. In the twisted case wrapping yields the twisted Heisenberg \textsc{xxx} chain, i.e. the chain with twisted boundary conditions, and a family of non-trivial twistings of the HS chain: the FK chains. 
\subsection{Periodic wrapping}
The infinite-length Heisenberg \textsc{xxx} spin chain is defined by the hamiltonian
\begin{equation}
	\label{eq:inf_Heis}
	H_{\textsc{xxx}} = \frac{1}{2} \sum_{i \in \Z} h^{\textsc{xxx}}_i\, ,\quad h^{\textsc{xxx}}_i = 1-P_{i,i+1}\, , 
\end{equation}
where $P_{ij}  = \frac{1}{2} (1 + 2\sigma_i^+ \sigma_{j}^- + 2\sigma_i^- \sigma_{j}^+ + \sigma_i^z \sigma_{j}^z)$ is the spin exchange operator acting on two sites $i$ and $j$ of the space $\ldots \otimes V \otimes  \ldots \otimes V \otimes \ldots$ with $V \cong \C \ket{\uparrow} \oplus \C \ket{\downarrow}$. Its building blocks, the Pauli spin matrices $\sigma_i^\alpha$ (with $\alpha = +,-,z$) attached to a fixed site $i$, furnish a $\mathfrak{sl}_2$-representation on each local vector space $V$. The summand  $h^{\textsc{xxx}}_i$ in \eqref{eq:inf_Heis} describes how the particle at site $i$ interacts with the other particles. 

Fix some length $N\in \mathbb{N}$ and define 
\begin{equation}
	h^{\textsc{xxx},N}_i = 1-P_{i,i \text{ mod } N}\, , 
\end{equation}
which 'wraps' the interaction onto a chain of finite length $N$. For this simple case only the boundary interaction described by $h^{\textsc{xxx},N}_N$ gets altered compared to the unwrapped case. Now upon imposing the periodic boundary conditions 
\begin{equation}
	\label{eq:periodic_boundary_conditions}
	\sigma_{i+N}^\alpha = \sigma_i^\alpha\, , \quad \alpha = +,-,z\, , \quad i \in \Z\, , 
\end{equation}
the sum over $N$ consecutive interaction summands yields the finite periodic Heisenberg hamiltonian
\begin{equation}
	\label{eq:fin_Heis}
		H_{\textsc{xxx}} = \frac{1}{2} \sum_{i=1}^{N-1} \big( 1-P_{i,i+1} \big) + \frac{1-P_{1,N}}{2}\, . 
\end{equation}

We can perform the same procedure on a long-range chain such as the rational spin-$1/2$ Haldane--Shastry chain, with hamiltonian
\begin{equation}
	\label{eq:rat_HS}
	H_\textsc{hs}^\text{rat} = \frac{1}{2} \sum_{i \in \Z} h_i\, , \,\, \text{ with } h_i \coloneqq \frac{1}{4}\sum_{\substack{ j \in \Z \\ j \neq i	 }} \frac{1-P_{ij}}{(i-j)^2}\, . \vspace{-10pt}
\end{equation}
Rewriting the $h_i$ as 
\begin{equation}
	\label{eq:hn_ham}
	h_i = \frac{1}{4}\sum_{j =1}^{N-1} \sum_{l\in \mathbb{Z}}\frac{1-P_{i,i+j+l N}}{(n+l N)^2}
\end{equation}
and again imposing the periodic boundary conditions \eqref{eq:periodic_boundary_conditions}, 
which imply $P_{i,j+N} = P_{ij}$, we can perform the sum over $l$ in \eqref{eq:hn_ham} explicitly to yield
\begin{equation}
	\label{eq:hn_periodic}
	h_i^{(N)} = \left(\frac{\pi}{N}\right)^2 \sum_{\substack{j=1 \\ j \neq i}}^{N-1}\frac{1-P_{i,i+j \text{ mod } N}}{4\sin^2 \tfrac{\pi}{N} (i-j)}\, . 
\end{equation}
Upon dropping the unimportant overall scaling and again summing over the first $N$ summands only, this defines the (trigonometric) Haldane-Shastry chain 
\begin{equation}
	\label{eq:HS_ham}
	H_\textsc{hs} = \frac{1}{2} \sum_{i=1}^N h_i^{(N)} =  
	%\left(\frac{\pi}{N}\right)^2
	 \sum_{i=1}^N \sum_{j>i}^N \frac{1-P_{ij}}{4 \sin^2 \tfrac{\pi}{N} (i-j)}\,
\end{equation}
acting on the finite-dimensional Hilbert space $V^{\otimes N}$. The chain has an $\mathfrak{sl}_2$-symmetry (i.e. is \emph{isotropic}), as the hamiltonian \eqref{eq:HS_ham} commutes with the action of  $\mathfrak{sl}_2$ on $V^{\otimes N}$ generated by 
\begin{equation}
	\begin{aligned}
		S^\pm &\coloneqq \sum_{i=1}^N \sigma^\pm_i\, , \qquad \,\,\,\, S^z \coloneqq \frac{1}{2}\sum_{i=1}^N \sigma^z\, , \\
		  [S^\pm, S^z] &\phantom{:}= \pm S^\pm\, , 
		\quad [S^+, S^-] = 2 S^z\, . 
	\end{aligned}
\end{equation}
Remarkably, the hamiltonian $H_\textsc{hs}$ commutes with three further  generators
\begin{equation}
	\label{eq:HS_yangian}
	\begin{aligned}
		Q^\pm &\coloneqq \mp \frac{1}{2} \sum_{i<j}^N \cot(\tfrac{\pi}{N}(i-j) (\sigma_i^\pm \sigma_j^z- \sigma_i^z\sigma^\pm_j)\, , \\		Q^z &\coloneqq \frac{1}{2} \sum_{i<j}^N \cot(\tfrac{\pi}{N}(i-j) (\sigma^+_i \sigma_j^-- \sigma_i^-\sigma^+_j)\, ,
		\end{aligned}
\end{equation}
which transform in the adjoint representation of $\mathfrak{sl}_2$, 
\begin{equation}
	\label{eq:adjoint}
	\begin{aligned}
				 [S^z,Q^\pm] &= \pm Q^\pm\, , \quad [S^\pm,Q^\mp] = \pm 2 Q^z\, , \\
		[S^\pm ,Q^z]&= \mp Q^\pm \, , 
			\end{aligned}
\end{equation}
 and obey the closure (Serre) relation
\begin{equation}
	[Q^z,[Q^+,Q^-]] = (Q^+S^--S^+ Q^-)S^z\, . 
\end{equation}
They extend the $\mathfrak{sl}_2$-action to a $Y(\mathfrak{sl}_2)$-action, where they act as the `level-$1$' generators. This action, which still commutes with $H_\textsc{hs}$, is what underlies the integrability of the Haldane--Shastry chain. 

The spectrum of $H_\textsc{hs}$ decomposes into irreducible representations of $Y(\mathfrak{sl}_2)$, which are characterised by a ($\mathfrak{gl}(2|0)$-)\emph{motif}: a sequence of increasing integers $\mu_i \in \{1,\ldots, N-1\}$ such that 
\begin{equation}
	\label{eq:motif_relation}
	\mu_i > \mu_{i-1}+1\, . 
\end{equation}
Motifs uniquely specify the Drinfel'd polynomial of a Yangian representation, but can also be related to the physics of the chain directly. The elementary excitations over the antiferromagnetic \footnote{This technically corresponds to the spectrum of $-H_\textsc{hs}$, which can straightforwardly be obtained from $H_\textsc{hs}$, and for simplicity we will continue to treat the ferromagnetic case.} ground-state, the spinons, exhibit fractional exchange statistics \cite{Hal_91a}. The spinons organise in orbital bands, and adding a new spinon reduces the number of available bands by $1/2$, implying they behave as a type of anyons known as semions \cite{Hal_94}. The motif constituents $\mu_i$ label the gaps between the different spinon orbitals.

The Yangian representation associated to $\mu$ from a Yangian highest weight vector $\ket{\mu} \in V^{\otimes N}$ satisfying
\begin{equation}
	\label{eq:Yangian_hw_vector}
	S^+ \ket{\mu} = 0  = Q^+ \ket{\mu}\, ,
\end{equation}
with momentum and energy expressed in terms of the motif via
\begin{equation}
	\label{eq:HS_eigenvalues}
	\begin{aligned}
	P &= \frac{2\pi}{N}\sum_i \mu_i\, , \quad E_\textsc{hs} = \sum_{i} \varepsilon(\mu_i)\, , \\
	 \varepsilon(\mu) &= \frac{1}{2}\mu(N-\mu)\, , \quad \lambda^z = \I \sum_i \left(\frac{N}{2}-\mu_i\right)\, , 
	 \end{aligned}
\end{equation}
with $\varepsilon$ the dispersion relation and $\lambda^z$ the $Q^z$-eigenvalue of $\ket{\mu}$. An explicit expression for $\ket{\mu}$ can be given in terms of Jack polynomials \cite{Hal_91a}. The other vectors in this representation -- the \emph{descendants} -- are generated by the repeated action of $S^-$ and $Q^-$ on $\ket{\mu}$. 

The (periodic, left) translation operator
\begin{equation}
	\label{eq:periodic_translation}
	T(0) \coloneqq P_{N-1,N}\cdots P_{12} = 	\tikz[baseline={([yshift=-.5*11pt*0.2]current bounding box.center)},xscale=0.4,yscale=0.2,font=\footnotesize]{
		\draw[rounded corners=2pt,->] (0,0) -- (0,1) -- (5,6) -- (5,7.5);
		\draw[rounded corners=2pt,->] (1,0)  -- (1,1) -- (0,2) -- (0,7.5);
		\foreach \x in {2,...,4} \draw[rounded corners=2pt,->] (\x,0) -- (\x,\x) -- (\x-1,\x+1) -- (\x-1,7.5);
		\foreach \x in {-1,...,1} \draw (3+.2*\x,-1.4);
		\foreach \x in {-1,...,1} \draw (2+.2*\x,9.4);
		\draw[rounded corners=2pt,->] (5,0) -- (5,5) -- (4,6) -- (4,7.5);
		%\draw[fill=black] (4.8,6) rectangle ++(0.4,0.9);
	}\, , 
\end{equation}
with the nearest-neighbour exchange operators depicted by
\begin{equation}
		 P_{i,i+1} = \tikz[baseline={([yshift=-.5*11pt*0.3]current bounding box.center)},xscale=0.4,yscale=0.2,font=\footnotesize]{
		\draw[->] (0,0) -- (0,3);
		\foreach \x in {-1,...,1} \draw (.75+.2*\x,1.5) node{$\cdot\mathstrut$};	
		\draw[->] (1.5,0) -- (1.5,3);
		\draw[rounded corners=2pt,->] (3.5,0) -- (3.5,1) -- (2.5,2) -- (2.5,3);
		\node[below] at (3.5,0) {\scalebox{0.8}{$i+1$}};
			\node[below] at (2.5,0) {\scalebox{0.8}{$i$}};
		\draw[rounded corners=2pt,->] (2.5,0) -- (2.5,1) -- (3.5,2) -- (3.5,3);
		\draw[->] (4.5,0) -- (4.5,3);
		\foreach \x in {-1,...,1} \draw (5.25+.2*\x,1.5) node{$\cdot\mathstrut$};	
		\draw[->] (6,0) -- (6,3);
		} \, ,
		\vspace{-5pt}
\end{equation}
commutes with  $H_{\textsc{hs}}$: $[H_{\textsc{hs}}, T(0)] = 0$. 

\subsection{Twisted wrapping}
Starting again from the infinite-length hamiltonian \eqref{eq:inf_Heis} we can similarly incorporate twisted boundary conditions
\begin{equation}
	\sigma_{i+l N}^\alpha = \E^{\I \pi l \varphi S^z}\sigma_i \E^{-\I \pi l \varphi S^z}\, , \quad \alpha \in \{ +,-,z \}\, , \quad l \in \Z\, , 
\end{equation}
where $\varphi \in \mathbb{Q}$ and $S^z$ is the global spin-$z$ operator. This twists the spins in the $x$- and $y$-directions 
\begin{equation}
	\sigma_{i+l N}^{\pm} = \E^{\pm \I \pi l \varphi } \sigma_{i}^\pm\, , \quad \sigma_{i+l N}^z = \sigma_i^z\, , 
\end{equation} 
and for the nearest-neighbour case only affects the boundary term, which becomes
\begin{equation}
	h_i^{\textsc{xxx},N} =  -\E^{ \I \pi \varphi} \sigma_N^+ \sigma_{1}^-  - \E^{- \I \pi \varphi}  \sigma_N^-  \sigma_{1}^+ + \frac{1-\sigma_N^z \sigma_1^z}{2} \, . 
\end{equation}
The resulting system is the well-known Heisenberg \textsc{xxx} chain with twisted boundary conditions:
\begin{equation}
	\begin{aligned}
	H_{\textsc{xxx}}^\text{tw} &= \frac{1}{2}\sum_{i=1}^{N-1} (1-P_{i,i+1})+ \frac{1-\sigma_N^z \sigma_1^z}{2}\\
	&\, -\frac{\E^{ \I \pi \varphi} \sigma_N^+ \sigma_{1}^-  + \E^{- \I \pi \varphi}  \sigma_N^-  \sigma_{1}^+}{2} \, .
	\end{aligned} 
\end{equation}
Fukui and Kawakami showed in \cite{fukui1996exact} that one can perform the same twisted wrapping procedure on the rational HS chain \eqref{eq:rat_HS}, yielding (up to the same global factor $(\pi/N)^2$ from \eqref{eq:hn_periodic})
\begin{align}\label{eq:HFKdef}
	H_{\textsc{fk}}(p/q)& \coloneqq  	\sum_{i<j}\bigg[ \frac{1}{4\sin^2\frac{\pi}{N}(i-j)} \frac{1-\sigma_i^z\sigma_j^z}{2}  \\
	&\,-\sum_{m=0}^{q-1}\frac{e^{2\pi \I m\frac{p}{q}}\sigma_i^+\sigma_j^-+e^{-2\pi \I m\frac{p}{q}}\sigma_i^-\sigma_j^+}{4 q^2\sin^2\frac{\pi}{qN}(i-j+mN)} \bigg]\, , \nonumber 
\end{align}
with $p\in \Z$ and $q\in \mathbb{N}$ and $p/q$ any representatives of $\varphi \in \mathbb{Q}$  \footnote{Compared to \cite{fukui1996exact} the hamiltonians in \eqref{eq:HFKdef} are shifted by an immaterial, constant matrix.}. 

\section{Properties}
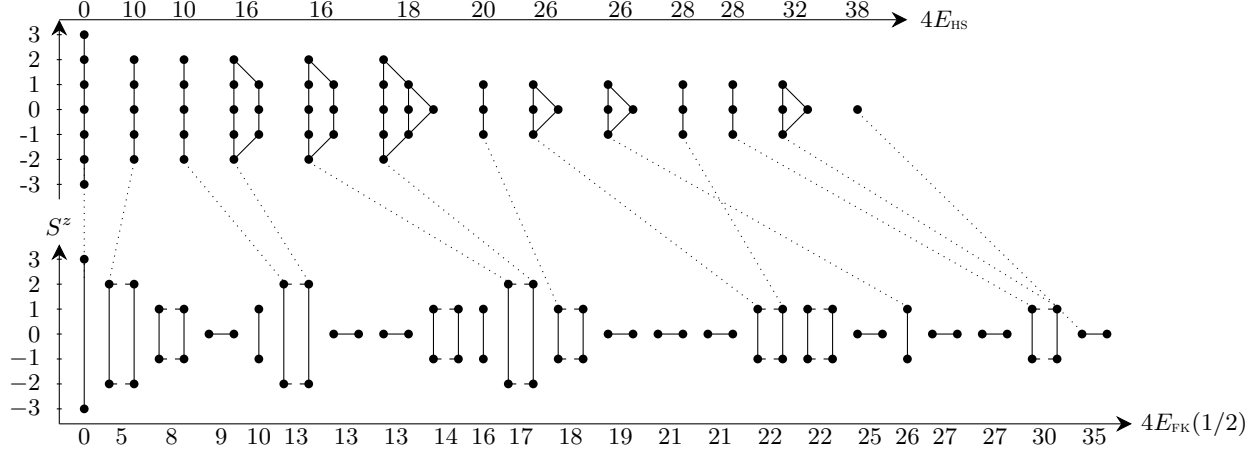
\begin{figure*}
	\begin{tikzpicture}[scale=0.33]
		\def\FKspace{9}
		% Draw y-axis on the left
		\draw[->, >={Stealth[length=6pt, width=6pt]}] (-1,-3.6) -- (-1,3.6);
		\foreach \y in {-3,...,3} {
			\draw (-1.1,\y) -- (-0.9,\y);
			\node[left] at (-1.4,\y) {\y};
		}
		%\node at (-2,4.3) {$S^z$};

		\draw[->, >={Stealth[length=6pt, width=6pt]}] (-1,3.6) -- (33,3.6);
		\node at (34.5,3.6) {$4 E_\textsc{hs}$};
		
		% First series (top) - centered at y=0
		\foreach \h/\x in {
			6/0,   4/2,   4/4,
			4/6,   2/7,          % connected pair
			4/9,   2/10,         % connected pair
			4/12,  2/13,  0/14,  % connected triple
			2/16,
			2/18,  0/19,         % connected pair
			2/21,  0/22,         % connected pair
			2/24,
			2/26,
			2/28,  0/29,         % connected pair
			0/31}
		{
			\pgfmathtruncatemacro{\hh}{\h/2}
			\ifnum\h>0
			\draw (\x,-\hh) -- (\x,\hh);
			\fi
			\foreach \y in {-\hh,...,\hh} {\fill (\x,\y) circle (5pt);}
		}
		% Connections for first series
		\draw (6,2)   -- (7,1);
		\draw (6,-2)  -- (7,-1);
		\draw (9,2)   -- (10,1);
		\draw (9,-2)  -- (10,-1);
		\draw (12,2)  -- (13,1) -- (14,0);
		\draw (12,-2) -- (13,-1) -- (14,0);
		\draw (18,1)  -- (19,0);
		\draw (18,-1) -- (19,0);
		\draw (21,1)  -- (22,0);
		\draw (21,-1) -- (22,0);
		\draw (28,1)  -- (29,0);
		\draw (28,-1) -- (29,0);
		% Energy labels above each representation
		\foreach \x/\E in {0/0, 2/10, 4/10, 6.5/16, 9.5/16, 13/18, 16/20, 18.5/26, 21.5/26, 24/28, 26/28, 28.5/32, 31/38} {
			\node at (\x,4) {\E};
		}
		
		% II. Connections between the 2 series
		\draw[dotted] (0,-2-0.1)  -- (0,-\FKspace+2+0.1);
		\draw[dotted] (2,-2-0.1)  -- (1,-\FKspace+2+0.1);
		\draw[dotted] (4,-2-0.1)  -- (8,-\FKspace+2+0.1);
		\draw[dotted] (6,-2-0.1)  -- (9,-\FKspace+2+0.1);
		\draw[dotted] (9,-2-0.1)  -- (17,-\FKspace+2+0.1);
		\draw[dotted] (12,-2-0.1)  -- (18,-\FKspace+2+0.1);
		
		\draw[dotted] (16,-1-0.1)  -- (19,-\FKspace+1+0.1);
		\draw[dotted] (18,-1-0.1)  -- (27,-\FKspace+1+0.1);
		\draw[dotted] (21,-1-0.1)  -- (33,-\FKspace+1+0.1);
		\draw[dotted] (24,-1-0.1)  -- (28,-\FKspace+1+0.1);
		\draw[dotted] (26,-1-0.1)  -- (38,-\FKspace+1+0.1);
		\draw[dotted] (28,-1-0.1)  -- (39,-\FKspace+1+0.1);
		\draw[dotted] (31,-0.1)  -- (40,-\FKspace+0.1);

		% III. Second series (bottom) centered at y=-\FKspace
		% Draw y-axis on the left
		\draw[->, >={Stealth[length=6pt, width=6pt]}] (-1,-\FKspace-3.6) -- (-1,-\FKspace+3.6);
		
		\draw[->, >={Stealth[length=6pt, width=6pt]}] (-1,-\FKspace-3.6) -- (42,-\FKspace-3.6);
		\node at (44.5,-\FKspace-3.6) {$4 E_\textsc{fk}(1/2)$};

		\pgfmathsetmacro{\ymin}{-3-\FKspace}
		\pgfmathsetmacro{\ymax}{3-\FKspace}
		\foreach \y in {\ymin,...,\ymax} {
			\draw (-1.1,\y) -- (-0.9,\y);
			\pgfmathsetmacro{\ylabel}{\y+\FKspace}
			\node[left] at (-1.4,\y) {\pgfmathprintnumber{\ylabel}};
		}
		\node at (-1,4.3-\FKspace) {$S^z$};
		
		% Vertical lines
		\foreach \h [count=\x from 0] in {6,4,4,2,2} {
			\ifnum\h>0
			\draw (\x,-\h/2-\FKspace) -- (\x,\h/2-\FKspace);
			\fill (\x,-\h/2-\FKspace) circle (5pt);
			\fill (\x,\h/2-\FKspace) circle (5pt);
			\fi
		}
		\foreach \h [count=\x from 7] in {2,4,4} {
			\ifnum\h>0
			\draw (\x,-\h/2-\FKspace) -- (\x,\h/2-\FKspace);
			\fill (\x,-\h/2-\FKspace) circle (5pt);
			\fill (\x,\h/2-\FKspace) circle (5pt);
			\fi
		}
		\foreach \h [count=\x from 14] in {2,2,2,4,4,2,2} {
			\ifnum\h>0
			\draw (\x,-\h/2-\FKspace) -- (\x,\h/2-\FKspace);
			\fill (\x,-\h/2-\FKspace) circle (5pt);
			\fill (\x,\h/2-\FKspace) circle (5pt);
			\fi
		}
		\foreach \h [count=\x from 27] in {2,2,2,2} {
			\ifnum\h>0
			\draw (\x,-\h/2-\FKspace) -- (\x,\h/2-\FKspace);
			\fill (\x,-\h/2-\FKspace) circle (5pt);
			\fill (\x,\h/2-\FKspace) circle (5pt);
			\fi
		}
		\foreach \h [count=\x from 33] in {2} {
			\ifnum\h>0
			\draw (\x,-\h/2-\FKspace) -- (\x,\h/2-\FKspace);
			\fill (\x,-\h/2-\FKspace) circle (5pt);
			\fill (\x,\h/2-\FKspace) circle (5pt);
			\fi
		}
		\foreach \h [count=\x from 38] in {2,2} {
			\ifnum\h>0
			\draw (\x,-\h/2-\FKspace) -- (\x,\h/2-\FKspace);
			\fill (\x,-\h/2-\FKspace) circle (5pt);
			\fill (\x,\h/2-\FKspace) circle (5pt);
			\fi
		}
		
		% Horizontal lines with dots at each integer
		\foreach \start/\end in {5/6, 10/11, 12/13, 21/22, 23/24, 25/26, 31/32, 34/35, 36/37, 40/41} {
			\draw (\start,-\FKspace) -- (\end,-\FKspace);
			\foreach \x in {\start, ..., \end} {\fill (\x,-\FKspace) circle (5pt);}
		}
		
		%Parity connections:
		\draw[dashed] (1,4/2-\FKspace) -- (2,4/2-\FKspace);
		\draw[dashed] (1,-4/2-\FKspace) -- (2,-4/2-\FKspace);
		
		\draw[dashed] (3,2/2-\FKspace) -- (4,2/2-\FKspace);
		\draw[dashed] (3,-2/2-\FKspace) -- (4,-2/2-\FKspace);
		
		\draw[dashed] (8,4/2-\FKspace) -- (9,4/2-\FKspace);
		\draw[dashed] (8,-4/2-\FKspace) -- (9,-4/2-\FKspace);
		
		\draw[dashed] (14,2/2-\FKspace) -- (15,2/2-\FKspace);
		\draw[dashed] (14,-2/2-\FKspace) -- (15,-2/2-\FKspace);
		
		\draw[dashed] (17,4/2-\FKspace) -- (18,4/2-\FKspace);
		\draw[dashed] (17,-4/2-\FKspace) -- (18,-4/2-\FKspace);
		
		\draw[dashed] (19,2/2-\FKspace) -- (20,2/2-\FKspace);
		\draw[dashed] (19,-2/2-\FKspace) -- (20,-2/2-\FKspace);
		
		\draw[dashed] (27,2/2-\FKspace) -- (28,2/2-\FKspace);
		\draw[dashed] (27,-2/2-\FKspace) -- (28,-2/2-\FKspace);
		
		\draw[dashed] (29,2/2-\FKspace) -- (30,2/2-\FKspace);
		\draw[dashed] (29,-2/2-\FKspace) -- (30,-2/2-\FKspace);
		
		\draw[dashed] (38,2/2-\FKspace) -- (39,2/2-\FKspace);
		\draw[dashed] (38,-2/2-\FKspace) -- (39,-2/2-\FKspace);
		
		% Energy labels below each representation
		\foreach \x/\E in {0/0, 1.5/5, 3.5/8, 5.5/9, 7/10, 8.5/13, 10.5/13, 12.5/13, 14.5/14, 16/16, 17.5/17, 19.5/18, 21.5/19, 23.5/21, 25.5/21, 27.5/22, 29.5/22, 31.5/25, 33/26, 34.5/27, 36.5/27, 38.5/30, 40.5/35} {
			\node[align=center] at (\x,-\FKspace-4.1) {\E};
		}
	\end{tikzpicture}
	\caption{Schematic picture of the spectra of the twisted and untwisted chains for length $N=6$. Each dot represents an eigenvector. The vectors in the upper picture represent the untwisted (HS) chain, and are grouped by Yangian representations and labelled by four times their energy. The lower picture represents the twisted chain (at $\varphi=1/2$), with the vectors grouped into quadruplets (or doublets) as generated by the action of the spin-flip and parity operators, also labelled by four times their energy. The dotted lines between the upper and lower spectra indicate how the Yangian-highest-weight state of a Yangian representation in the untwisted chain gives rise to an eigenvector in the twisted chain, through the identity \eqref{eq:towards_FK_eigenvectors}.}
	\label{fg:FKvsHS_chain_picture}
\end{figure*}

The hamiltonians \eqref{eq:HFKdef} are hermitian. Note that choosing $\varphi \in \Z$ returns us to the untwisted Haldane--Shastry chain \eqref{eq:HS_ham}. Since $H_\textsc{fk}(\varphi +1) = H_\textsc{fk}(\varphi)$ we can restrict $\varphi \in [0,1]$. In this range the change of variable $\varphi \to 1-\varphi$ corresponds to a particle-hole transformation of the chain ($\ket{\uparrow} \leftrightarrow \ket{\downarrow}$), which can be realised by the spin-flip operator 
\begin{equation}
	\label{eq:spinflip}
	S_{\textsc{sf}} = \prod_{i=1}^N \sigma_i^x\, ,
\end{equation}
as follows: for $\varphi \in [0,1]$ 
\begin{equation}
	H_\textsc{fk}(\varphi) = S_{\textsc{sf}}  H_\textsc{fk}(1-\varphi) S_{\textsc{sf}} \, . 
\end{equation}
We can thus restrict further such that $\varphi$ is taken from the set $\mathbb{Q} \cap [0,1/2]$. Finally, it is easy to see that neither a global spin flip by \eqref{eq:spinflip} nor a parity transformation $X$ (sending site $i \mapsto N-i+1$) are symmetries, but both satisfy
\begin{equation}
	\label{eq:spinflip_parity}
	S_{\textsc{sf}} H_\textsc{fk}(\varphi) S_{\textsc{sf}} = H_\textsc{fk}(-\varphi) = X H_\textsc{fk}(\varphi) X\, , 
\end{equation}
such that their product is in fact a (CP-like) symmetry.  Furthermore $H_\textsc{fk}(\varphi)^* = H_\textsc{fk}(-\varphi)$ (with $^*$ indicating complex conjugation), so by \eqref{eq:spinflip_parity} the action of the spin-flip and parity operators on an eigenvector $\vect{v}$ with (real) energy $E$ generically generate a quadruplet $\{\vect{v},S_{\textsc{sf}} \vect{v}^*,X\vect{v}^*,X S_{\textsc{sf}} \vect{v} \}$, see Fig.~\ref{fg:FKvsHS_chain_picture}  \footnote{Compare with Fig. 1 in \cite{lamers2022spin}, where the (Yangian) representations are ordered according to their momenta. To avoid too many crossing lines we have ordered the eigenvectors based on their energy}, with the only exceptions occurring when $\vect{v}$ is a parity- or spin-flip singlet, i.e. satisfies $S_{\textsc{sf}} \vect{v} = \vect{v}$ or $X \vect{v} = \vect{v}$.

The twisting breaks the global $\mathfrak{sl}(2)$-symmetry, such that only a global $S^z$-symmetry survives. Nevertheless, Fukui and Kawakami showed that one can twist the Yangian generator $Q^z$ \eqref{eq:HS_yangian} that commutes with the HS hamiltonian to obtain an extra charge
\begin{equation}\label{eq:QzFKdef}
	\begin{aligned}
	Q^z(p/q) &\coloneqq \frac{1}{2q}\sum_{i<j}\sum_{m=0}^{q-1}\cot\left[\frac{\pi}{qN}(i-j+mN)\right]\times \\
	&\left(e^{2\pi imp/q}\sigma_i^+\sigma_j^--e^{-2\pi imp/q}\sigma_i^-\sigma_j^+\right)\, , 
	\end{aligned}
\end{equation} 
which commutes with $H_\textsc{fk}(p/q)$. The two operators $S^z$ and $Q^z(p/q)$ commute, and generate a representation of $Y(\mathfrak{gl}_1)$ \cite{fukui1996exact}. 

The twisting also breaks the translational symmetry generated by \eqref{eq:periodic_translation}, but the latter can be modified. Noting that 
\begin{equation}
	\E^{-\I \pi \varphi \, \sigma^z} =
	\begin{pmatrix}
		e^{-\I\pi \varphi } & 0\\
		0 & e^{\I\pi \varphi }
	\end{pmatrix}
\end{equation}
one can write the translation operator for $H_{\textsc{fk}}$ as\vspace{-15pt}
\begin{equation}\label{eq:translationpq}
	T(\varphi) \coloneqq  \,\,
	\tikz[baseline={([yshift=-3.5*11pt*0.2]current bounding box.center)},xscale=0.4,yscale=0.2,font=\footnotesize]{
		\draw[rounded corners=2pt,->] (0,0) -- (0,1) -- (5,6) -- (5,7.5);
		\draw[rounded corners=2pt,->] (1,0)  -- (1,1) -- (0,2) -- (0,7.5);
		\foreach \x in {2,...,4} \draw[rounded corners=2pt,->] (\x,0) -- (\x,\x) -- (\x-1,\x+1) -- (\x-1,7.5);
		\foreach \x in {-1,...,1} \draw (2+.2*\x,9.4);
		\draw[rounded corners=2pt,->] (5,0) -- (5,5) -- (4,6) -- (4,7.5);
		\draw[fill=black] (4.8,6) rectangle ++(0.4,0.9);
	}\, , \quad \text{ with } \E^{\I \pi \varphi \, \sigma^z} = 
	\tikz[baseline={([yshift=-1.5*11pt*0.2]current bounding box.center)},xscale=0.4,yscale=0.2,font=\footnotesize]{
		\draw[rounded corners=2pt,->] (0,0) --(0,3);
		\draw[fill=black] (-0.2,1) rectangle ++(0.4,0.9);
	}\, , 
\end{equation}
which as a formula is $T(\varphi) = \E^{-\I \pi \varphi\, \sigma^z_N} T(0)$. This expression is not unique, as one can split the exponent $\E^{-\I \pi \varphi \, \sigma^z_N} = \prod_{j=1}^N \E^{-\I \pi \varphi \, \sigma^z_N/N}$ and use commutation relations to relocate each factor, for example such that each local space receives one such factor, yielding the alternative expression
\begin{equation}
	T(\varphi) = \E^{-\I \pi \varphi \, \sigma^z_N/N}\ordprod_{j=1}^{N-1} \E^{-\I \pi \varphi \, \sigma^z_j/N} P_{j,j+1}\, . 
\end{equation}
Rotating all particles once around the chain yields 
\begin{equation}
	T(\varphi)^N = \E^{- 2\pi  \I \varphi \, S^z}\, . 
\end{equation}
Effectively, such a rotation counts the number of magnon excitations of a state and adds a fractional phase for each, causing the periodicity breaking. We denote by
\begin{equation}
	\label{eq:translation_N}
 \bar{T}(\varphi) \coloneqq \E^{ 2\pi  \I \varphi  \, S^z/N} T(\varphi) \, , \qquad 	\bar{T}(\varphi)^N =~\mathbb{I}\, . 
\end{equation} 
the normalised translation operator. 

The above implies that the spectrum can be labelled by the $S^z$- and $\bar{T}(\varphi)$-eigenvalues (i.e.\ by magnon numbers and momenta, respectively).

\section{Connecting twisted to untwisted}
\label{sec:twisted_and_untwisted}
Fukui and Kawakami noted in \cite[eqn.~(6.3)]{fukuiSUvGeneralizationTwisted1997} that the conformal limit of the twisted hamiltonian \eqref{eq:HFKdef} can be obtained from that of the (untwisted) HS hamiltonian by adding a suitable multiple of the extra Yangian charge. As we prove here, the same is true for the finite chain \footnote{We thank Andrii Liashyk for showing us his joint result with Ivan Sechin for the case $\varphi=1/2$ of the relation \eqref{eq:HFKHHSrel}.}. Indeed, we find that
\begin{equation}\label{eq:HFK+QFK}
	\begin{aligned}
& H_{\textsc{fk}}+\I \varphi\, Q^z_{\textsc{fk}}=-\frac{1}{2}\sum_{i<j}\frac{1}{4\sin^2\frac{\pi}{N}(i-j)}\times \\
	&\Big(2e^{\frac{-2\pi \I \varphi }{N}(i-j)}\sigma_i^+\sigma_j^-+2e^{\frac{2\pi \I \varphi}{N}(i-j)}\sigma_i^-\sigma_j^++\sigma_i^z\sigma_j^z-1\Big)\, ,
	\end{aligned} 
\end{equation}
which is already quite close to \eqref{eq:HS_ham}. To obtain \eqref{eq:HFK+QFK} we simplify the left-hand side using the relation 
\begin{equation}\label{eq:HFK+QFKbig}
	\begin{aligned}
	&\frac{e^{\frac{-2\pi \I p }{q N}(i-j)}}{\sin^2\frac{\pi}{N}(i-j)}
	= \\
	&\frac{1}{q^2}\sum_{m=0}^{q-1} e^{2\pi \I \frac{p}{q}m} \Bigg(\frac{1}{\sin^2\left(\frac{\pi}{qN}(i-j+mN)\right)} \\
	&- 2 \I p \cot\left(\frac{\pi}{qN}(i-j+mN)\right)\Bigg)\, , 
	\end{aligned}
\end{equation}
which follows from the identity
\begin{equation}\label{eq:cottwisted}
	\frac{1}{q}\sum_{m=0}^{q-1} e^{2\pi \I \frac{p}{q}(z+m)}\cot\frac{\pi}{q}(z+m) = \frac{e^{\pi\I z}}{\sin\pi z}\, . 
\end{equation}
We prove this identity and how it implies \eqref{eq:HFK+QFKbig} in Appendix \ref{app:trig_identity}. 

To simplify \eqref{eq:HFK+QFK} we perform a further change of basis defined by
\begin{equation}
	U_{\textsc{fk}}=\prod_{j=1}^N e^{\I\pi \varphi \sigma_j^z j /N} \, , 
\end{equation}
which twists each local vector space by a multiple of $\varphi$ depending on the location in the chain. From the action of $U_{FK}$ on the spin operators
\begin{equation}
	\label{eq:UFK_relations}
		U_{\textsc{fk}}\sigma_j^\pm U_{\textsc{fk}}^{-1} = e^{ \pm \frac{2\pi \I \varphi}{N}j}\sigma_j^\pm  
\end{equation}
we can find that 
\begin{equation}
	\label{eq:U_on_translation}
	U_{\textsc{fk}}\bar{T}(\varphi)U_{\textsc{fk}}^{-1} = \bar{T}(0)\, . 
\end{equation}
Moreover, a direct computation shows that 
\begin{equation}\label{eq:HFKHHSrel}
	U_{\textsc{fk}}(H_{\textsc{fk}}+\I \varphi Q^z_{\textsc{fk}})U^{-1}_{\textsc{fk}}=H_{\textsc{hs}}\, . 
\end{equation}

This relation provides a recipe to construct eigenvectors for $H_\textsc{fk}$ from those of $H_{\textsc{HS}}$. However, (1) $H_{\textsc{HS}}$ (and hence also $H_{\textsc{fk}}+\I \varphi Q^z_{\textsc{fk}}$) has a very degenerate spectrum due to its $Y(\mathfrak{gl}(2))$-symmetry and has no canonical orthogonal basis in each eigenspace beyond its highest-weight states, and (2) it is non-trivial to determine whether any such basis is also a basis of eigenvectors for the (smaller) eigenspaces of $H_\textsc{fk}$. Conversely, if we determined a full orthogonal basis of simultaneous eigenvectors for $H_{\textsc{fk}}$ and $Q^z_{\textsc{fk}}$ this would simultaneously serve as a distinguished basis for $H_\textsc{hs}$, mimicking results for the Heisenberg \textsc{xxx} chain (where finding an eigenbasis for the twisted chain can be used to determine a distinguished eigenbasis for the untwisted chain). We believe the construction of such a basis would naturally connect to the twisted transfer matrix approach of \cite{jiangNormsOverlapsYangian2026}, and might give meaning to the auxiliary twist parameter used therein. 

In this work, we focus on the Yangian highest weight vectors and aim to make the connection between eigenvectors and eigenvalues explicit. We first note that using the relations \eqref{eq:UFK_relations} it follows that 
\begin{equation}
	U_{\textsc{fk}}Q_{\textsc{fk}}^zU_{\textsc{fk}}^{-1}=\,Q^z+\frac{\I}{2}(S^z+S^-S^+-\frac{N}{2})\, , 
\end{equation}
with $Q^z$ the untwisted level-1 charge defined in \eqref{eq:HS_yangian}. For a Yangian highest weight vector satisfying 
\begin{equation}
	\label{eq:hw_for_eigenvector}
	\begin{aligned}
	H_{\textsc{hs}} \ket{\mu} &= E_{\text{HS}} \ket{\mu}\, , \quad 	S^z \ket{\mu} = \left(\frac{N}{2}-M\right)\ket{\mu}\, , \\	
	Q^z \ket{\mu} &= \lambda^z \ket{\mu}\, , 
	  \quad \,\,\,\hspace{1.2pt} S^+ \ket{\mu} = 0\, ,
	\end{aligned}
\end{equation}
with $M \geq 0$ the integer magnon number, the relation \eqref{eq:HFKHHSrel} now implies that 
\begin{equation}
	\label{eq:towards_FK_eigenvectors}
	\begin{aligned}
&H_{\textsc{fk}}U^{-1}_{\textsc{fk}}\ket{\mu}=	U_{\textsc{fk}}^{-1}\left( H_{\textsc{hs}}-\I \varphi U_{\textsc{fk}}Q^z_{\textsc{fk}}U^{-1}_{\textsc{fk}}\right)\ket{\mu}\\
	&=	U_{\textsc{fk}}^{-1}\left( H_{\textsc{hs}}-\I \varphi (Q^z+\frac{\I}{2}(S^z+S^-S^+-\frac{N}{2}))\right)\ket{\mu}\\
	&=	\left( E_{\text{HS}}- \varphi (\I \lambda^z +  M/2)\right)U_{\textsc{fk}}^{-1}\ket{\mu}\, . 
\end{aligned}
\end{equation}
Hence such vectors are $H_{\textsc{fk}}$-eigenvectors, with the same momentum and magnon number as the HS state $\ket{\mu}$, and with their energy related to the HS energy via
\begin{equation}
	\label{eq:FK_energy_1}
E_{\textsc{fk}} = E_{\textsc{hs}}-\varphi(\I\lambda^z+M/2)\, . 
\end{equation}
Using the relation \eqref{eq:HS_eigenvalues} between the motif and the relevant eigenvalues we can simplify \eqref{eq:FK_energy_1} further to 
\begin{equation}
	\begin{aligned}
			E_{\textsc{fk}} &= \sum_i \varepsilon_\varphi(\mu_i)\, , \quad \text{with } \\
			 \varepsilon_\varphi(\mu) &= \frac{1}{2}(\mu_i+\varphi) (N-\mu_i-\varphi) +\frac{\varphi(\varphi-1)}{2} \, , 
	\end{aligned}
\end{equation}
with $\varepsilon_\varphi$ the twisted dispersion from \cite{fukui1996exact}. Written in this form, we recognise the shifted motif $\mu+\varphi$ and an offset $\varphi(\varphi-1)/2$ which makes it so that $\varepsilon_\varphi(0) =\varphi(N-1)/2$ but vanishes as $\varphi \to 0$. 

Due to \eqref{eq:U_on_translation} the quasimomenta of $U_{\textsc{fk}}^{-1}\ket{\mu}$ coincide (up to scaling) with the motifs, $k_i = \frac{2\pi \mu_i}{N}$, and hence can be encoded in Haldane's `Bethe-ansatz-like equations': for each $i=1,\ldots M$ 
\begin{equation}
	\label{eq:ABA_equations}
	k_i N = 2\pi I_i - (M-1)\pi + \pi\sum_{j \neq i} \mathrm{sign}(k_i -k_j)\, ,
\end{equation}
for some choice of non-coinciding mode numbers $I_i \in \Z_N$. 
This means our derivation has recovered the asymptotic Bethe ansatz equation (14) in \cite{fukui1996exact} \footnote{Note that therein the equation is written in terms of shifted momenta $\tilde{k} = k+\varphi$. Writing out the shift explicitly one recovers \eqref{eq:ABA_equations} up to a normalisation of the mode numbers.}. These equations can be solved explicitly to yield 
\begin{equation}
	\mu_j = I_j -M +j\, . 
\end{equation}
For mode numbers satisfying $M \leq I_1 < I_2 <\ldots <I_M \leq N-1$ this yields precisely all Yangian highest weight states labelled by a motif $\mu$, of which there are 
\begin{equation}
	\sum_{M=0}^{\lfloor N/2 \rfloor} \binom{N-M}{M} = \textrm{Fib}(N+1)\, , 
\end{equation}
the $N+1$st Fibonacci number. This nicely matches the known number of motifs satisfying \eqref{eq:motif_relation}.

All remaining states are deformations of states which for the HS chain are descendants, and have the same energy and momentum as the Yangian highest weight states, but other data such as explicit eigenvectors requires separate analysis, see e.g. \cite{jiangNormsOverlapsYangian2026}. By a slight abuse of nomenclature we will similarly call the FK-deformations of such states \emph{deformed descendants}.

If we allow for the first few $I_i$ to equal $I_i = M-i$, we obtain motifs with repeated zeros, and parametrise some of the HS-descendants for which some of the quasimomenta vanish, obtained by acting with $Q^-$ or $S^-$ on $\ket{\mu}$. In line with \eqref{eq:FK_energy_1} such motifs also describe energies in the FK spectrum. Since this still excludes some sets of mode numbers $I_i \in \Z_N$, a simple counting shows that this does not cover all $2^N$ independent states. As we will see below, these states are not covered by other choices of integer mode numbers. For deformations of the other descendants the computation in \eqref{eq:towards_FK_eigenvectors} becomes unwieldy: working out the $Q^z$ and $S^-S^+$ contributions yields commutation relations of $Q^-$ with $Q^z$, generating a level-$2$ element of the Yangian. Thus, for such states we do not have a direct analogue of \eqref{eq:FK_energy_1}.

\section{Low-lying spectrum}
To explore which states cannot be covered by \eqref{eq:ABA_equations} we consider the first few magnon sectors. 

\begin{figure*}
{ \centering
\begin{tikzpicture}
	\def\x{4};
	\def\y{2.7};
\node at (-\x,\y) {		
		\includegraphics[width=8cm]{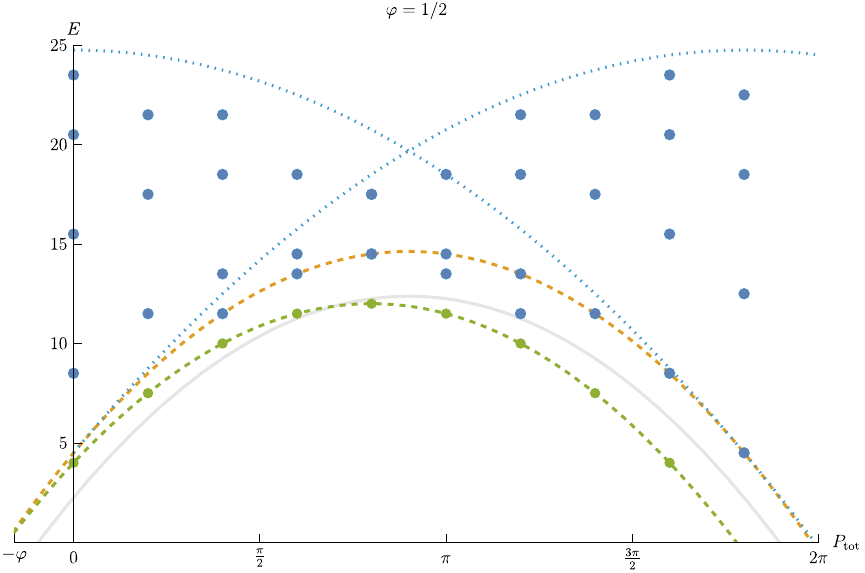}
				};
	
\node at (\x,\y) {		
	\includegraphics[width=8cm]{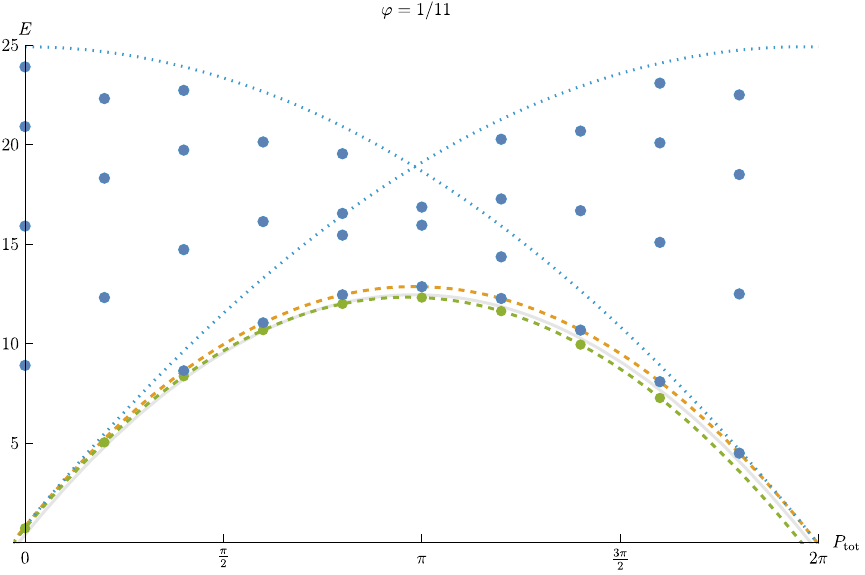}
	};

\fill [white] (-8,0.1) rectangle (-7.75,0.3);
\node at (-7.85,0.18) {\scalebox{0.5}{$\tilde{\varphi}$}};
\node at (0.1,0.18) {\scalebox{0.5}{$\tilde{\varphi}$}};

\end{tikzpicture}
}
	\caption{The two magnon spectrum at length $N=10$ for $\varphi=1/2$ and $\varphi=1/11$ in the energy-momentum plane, with $\tilde{\varphi} = 2\pi 
	\varphi/N$. The blue dots are motif states as parametrised by \eqref{eq:ABA_equations} (allowing for $\mu_1=0$), with energies that are sums of single particle energies, and lie above the single particle dispersion (indicated in gray) but below the (two-valued, blue dotted) curve of two identical magnons. The orange dashed line contains the descendants obtained by adding a zero-momentum excitation. The green dashed line contains the descendants (as green dots) which cannot be obtained from \eqref{eq:ABA_equations}. In the limit $\varphi \to 0$ the three curves, as well as the dots on them, coincide (see e.g. Fig.12 in \cite{klabbers2022coordinate}), in accordance with the high degeneracy of the Yangian symmetry of the untwisted chain.}
	\label{fg:M2_sector}
\end{figure*}
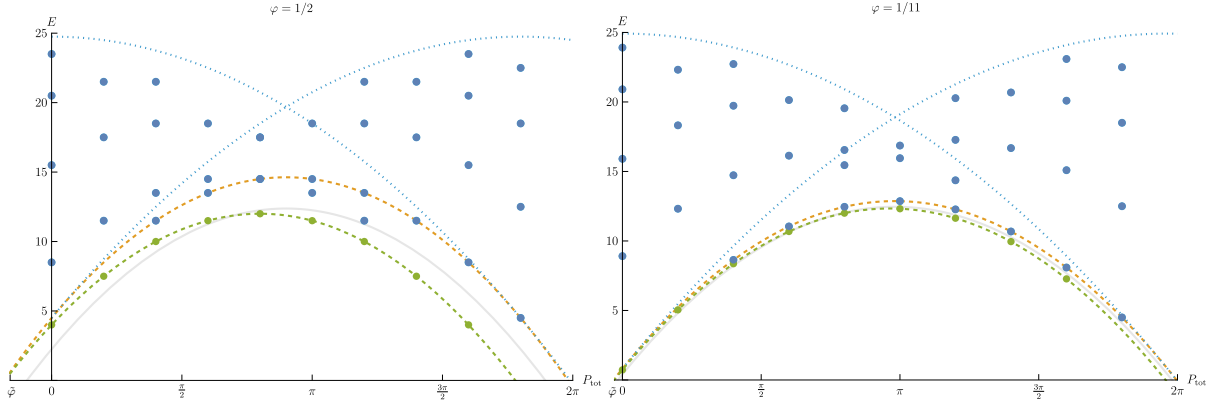
The sector with $M=0$ is spanned by $\ket{\uparrow\ldots \uparrow} = \ket{\emptyset}$, the state corresponding to the empty motif. It has energy $E(\varphi)=0$. The extra charge $Q^z(\varphi)$ vanishes on this sector, while $\bar{T}(\varphi)$ acts as the identity. 

In the one-magnon sector with $M=1$ the eigenvectors of \eqref{eq:HFKdef} are fixed by translational invariance. Given any single one-magnon `starting' vector $\ket{\vect{v}}$ we can write such eigenvectors as
\begin{equation}
	\sum_{n=1}^N \E^{2\pi \I n \mu/N} \bar{T}(\varphi)^{1-n} \ket{\vect{v}}\, , 
\end{equation}
with $\mu\in \{0,1,\ldots, N-1\}$. As we will see shortly, it makes sense to choose $\ket{\vect{v}} = U_\textsc{fk}^{-1}\ket{1}$ in terms of one of the coordinate basis vectors $\ket{n} \coloneqq \sigma_n^- \ket{\uparrow\ldots \uparrow} =  \ket{\uparrow \ldots \downarrow \ldots \uparrow}$ with the $\downarrow$ in the $n$th position. One checks straightforwardly that this vector has $\bar{T}(\varphi)$-eigenvalue $\E^{2 \pi \I \mu/N}$, consistent with \eqref{eq:translation_N}, and energy $\varepsilon_\varphi(\mu)$. The non-zero $\mu$ can be associated to motifs: the $N$ motifs $\{\mu\}$ define one-magnon eigenvectors for the HS-chain by
\begin{equation}
	\ket{\{\mu\}} = \sum_{n=1}^N \E^{2\pi \I n \mu/N} \ket{n}\, .
\end{equation} 
Noting that $\ket{n} = \bar{T}(0)^{1-n} \ket{1}$ and utilising the basis transformation $U_\textsc{fk}$ with \eqref{eq:U_on_translation} we find that 
\begin{equation}
	U_\textsc{fk}^{-1} \ket{\{\mu\}} = \sum_{n=1}^N \E^{2\pi \I n \mu/N} \bar{T}(\varphi)^{1-n} U_\textsc{fk}^{-1}\ket{1}
\end{equation}
is an $H_\textsc{fk}$-eigenvector, providing the rationale for setting $\ket{\vect{v}} = U_\textsc{fk}^{-1}\ket{1}$. The state with $\mu=0$ is a deformed descendant of $\ket{\emptyset}$, and can be described by adding a zero to its motif: $\emptyset \cup \{0\} = \{0\}$. Its energy is $\varepsilon_\varphi(0) \neq 0$ for $\varphi \neq 0$. This yields all $\binom{N}{1} = N$ eigenvectors in the one-magnon sector. 

The two-magnon sector (see Fig.\ \ref{fg:M2_sector}) contains motif-states analogous to the one-magnon case: a motif $\mu = \{\mu_1, \mu_2\}$ defines a Yangian highest weight state $\ket{\mu}$ such that $U_\textsc{fk}^{-1} \ket{\mu}$ is an eigenvector of $H_\textsc{fk}$ with eigenvalue $\varepsilon_\varphi(\mu_1) + \varepsilon_\varphi(\mu_2)$ (the blue dots in Fig.~\ref{fg:M2_sector}). This accounts for $\binom{N-2}{2}$ -- the number of motifs of length $2$ -- of the $\binom{N}{2}$ two-magnon states. To find the remaining $2N-3$ states we study the deformed descendants of the one-magnon states numerically, i.e. for a wide range of twists and lengths. For the undeformed HS chain there are $N-3$ affine descendants obtained by acting with $Q^-$ -- and orthogonalise by subtracting a suitable multiple of the $S^-$ descendants-- on the $M=1$ highest weight states with motif $2\leq \mu \leq N-2$. It furthermore has $N$ $\mathfrak{sl}_2$-descendants, obtained by acting with $S^-$. Of these, $N-1$ are obtained from the $M=1$ highest weight states (with $1\leq \mu \leq N-1$) and $1$ from the $M=0$ ground state. 

For the twisted chain we find the following: for each $2\leq \mu\leq N-1$ we can find a descendant with energy $\varepsilon_\varphi(\mu) + \varepsilon_\varphi(0)$, which lies on the orange descendant-curve in Fig.~\ref{fg:M2_sector}, and corresponds to adding a zero-momentum excitation to a one-particle state. This excludes the lowest two values $\mu=0$ and $\mu=1$, in line with the motif condition \eqref{eq:motif_relation}, but includes $\mu = N-1$. We speculate that the state with $\mu=N-1$ becomes a pure $\mathfrak{sl}_2$-descendant in the untwisted limit, which is corroborated by numerical analysis. Neatly, these deformed descendants are naturally incorporated in the solutions of the ABA equations \eqref{eq:ABA_equations} by allowing $\mu_1$ to be equal to zero without abandoning the motif condition \eqref{eq:motif_relation}, i.e. by letting the mode numbers satisfy 
\begin{equation}
	M-1 \leq I_1 < I_2 <\ldots <I_M \leq N-1\, . 
\end{equation}
Together these account for $N-2$ of the deformed descendant states. We furthermore observe numerically that the remaining $N-1$ states lie on the green curve in Fig.~\ref{fg:M2_sector}, which can be parametrised as
\begin{equation}
	\label{eq:sec_desc}
	\varepsilon_\varphi(\mu+\varphi) + \frac{3}{2}(\varphi-1)\varphi = \varepsilon_\varphi(\mu+\varphi) + 3 \varepsilon_\varphi(-\varphi)\, . 
\end{equation}
The energies correspond to choosing $\mu \in \{0,\ldots,N-2\}$. By a direct check we note that for any choice of integer mode numbers compatible with the total momentum there are no solutions of \eqref{eq:ABA_equations} (not even complex ones \footnote{This requires a prescription for the extension of the $\mathrm{sign}$-function in \eqref{eq:ABA_equations} into the complex plane. We choose $\mathrm{sign}_\C(k) =\mathrm{sign}(\mathrm{Re}(k))$. A better choice could be to consider the trigonometric limit of the Bethe-like equations of the elliptic generalisation of the FK chains. We will do so in \cite{Thomas}.}) such that $\varepsilon_\varphi(\mu_1) + \varepsilon_\varphi(\mu_2)$ yields the energies of these deformed descendant states. Hence these $N-1$ deformed descendants are not captured by the equations \eqref{eq:ABA_equations}. 

Following the state with $\mu=0$ for decreasing $\varphi$ (see Fig.~\ref{fg:M2_sector}) one sees that it is the deformation of the double $\mathfrak{sl}_2$-descendant of the HS ground state, but it has energy $\varepsilon_\varphi(\varphi)+ \varepsilon_\varphi(-\varphi) \neq 2\varepsilon_\varphi(0)$, indicating that the deformed descendant mechanism is more complicated than for the undeformed chain. The second expression in \eqref{eq:sec_desc} shows that these energies can be written as an explicit sum of single-particle energies of excitations whose momenta sum up to the total momentum $2\pi \mu/N$, with the second term belonging to a different excitation with energy contribution $3 \varepsilon_\varphi(-\mu)$. This is highly non-trivial, leading us to conjecture there is a hidden structure which can explain the regularity of the spectrum. We defer the analysis of this structure and the spectrum for higher magnon numbers to future work. 
\newpage

\section{A tale of four spin chains}
\label{sec:tale}
Among all values for $\varphi$, the choice $\varphi=1/2$ is special. Inspecting the various symmetries we see that $H_\textsc{fk}(1/2)$ commutes with both the spin-flip and the parity operator and is manifestly real, providing some hints that there might be further hidden symmetry. In contrast, its two magnon spectrum in Fig.~\ref{fg:M2_sector} does not look distinct from any other value of $\varphi$. As we will see in this section, this model has appeared in various guises in the literature. 

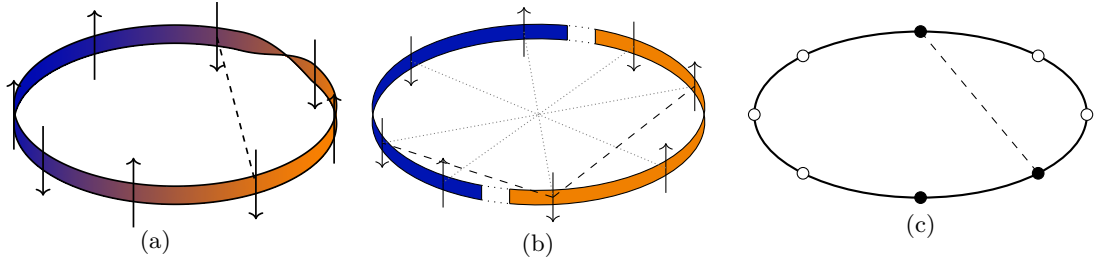
\begin{figure*}
	\centering
	\begin{minipage}{.3\textwidth}
		\centering
\includegraphics{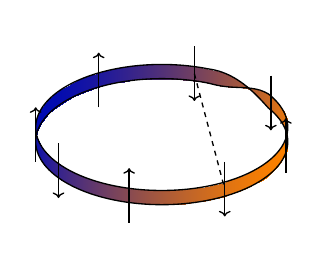} \\
\vspace{-25pt}
(a) \\
\vspace{25pt}
	\end{minipage}
	\begin{minipage}{.3\textwidth}
		\centering
	\includegraphics{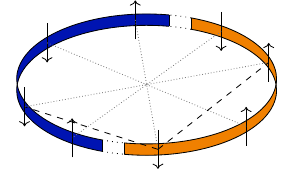} \\
	(b)
	\end{minipage}
	\begin{minipage}{.3\textwidth}
		\centering
		\begin{tikzpicture}[scale=1.1]
		\def\R{2}      % Radius of the chain
		\def\r{0.07}    % Radius of the small spin circles
		\def\angle{60} % Vertical viewing angle in degrees
		%The ellipse (circle viewed from vertical 45° angle)
		\draw[thick] (0,0) ellipse [x radius=\R, y radius={\R*cos(\angle)}];
		%Circles
		\foreach \i in {0,1,3,4,5} {
			\pgfmathsetmacro{\theta}{45*\i}
			\pgfmathsetmacro{\x}{\R*cos(\theta)}
			\pgfmathsetmacro{\y}{\R*sin(\theta)*cos(\angle)}
			\draw[fill=white] (\x,\y) circle (\r);
		}
		%Filled circles
		\foreach \i in {2,6,7} {
			\pgfmathsetmacro{\theta}{45*\i}
			\pgfmathsetmacro{\x}{\R*cos(\theta)}
			\pgfmathsetmacro{\y}{\R*sin(\theta)*cos(\angle)}
			\draw[fill=black] (\x,\y) circle (\r);
		}
		%Interaction dashed line between spins pair
    	\pgfmathsetmacro{\thetaA}{45*2}
    	\pgfmathsetmacro{\xA}{\R*cos(\thetaA)}
    	\pgfmathsetmacro{\yA}{\R*sin(\thetaA)*cos(\angle)}
    	\pgfmathsetmacro{\thetaB}{45*7}
    	\pgfmathsetmacro{\xB}{\R*cos(\thetaB)}
    	\pgfmathsetmacro{\yB}{\R*sin(\thetaB)*cos(\angle)}
    	\draw[dashed] (\xA,\yA) -- (\xB,\yB);
		\end{tikzpicture}
		\\
		(c)
	\end{minipage}
	\\
	\vspace{-25pt}
	\caption{Graphical depictions of three of the spin chains discussed in \textsection\ref{sec:tale}. In (a), the Fukui-Kawakami chain corresponding to $\varphi=1/2$, characterised by the antiperiodic boundary condition, with a typical pair interaction indicated by the dashed line. In (b), the chain introduced in \cite{basu-mallickNovelClassTranslationally2020}, which can be thought of as taking a chain (in orange) and gluing its spin-flipped mirror to it (in blue), with spins connected to their mirror by a dotted line. Considering interactions between each pair of spins, and between each pair of a spin and a mirror spin, as indicated by the dashed lines. In (c), the $\mathfrak{gl}(1|1)$ HS chain, with each site either empty ($\circ$) or occupied by a fermion ($\bullet$), with pair interactions between any pair of sites.}
	\label{fg:three_chains}
\end{figure*}

Novel chains can be constructed out of the HS chain in other ways beyond wrapping or twisting. For example, take a chain of length $N$, create a mirror copy by spin-flipping it and glue the two chains together into a closed chain. Define a hamiltonian as acting with the Haldane--Shastry pair interaction on each pair of original spins and on each pair of an original and a mirror copy spin (see Fig.\ref{fg:three_chains}). 
This model was introduced in \cite{basu-mallickNovelClassTranslationally2020} as the minimally polarised version of a family of chains, and has hamiltonian
\begin{equation}
	\label{eq:HBM}
	\begin{aligned}
	H_{\textsc{bmfgl}} \coloneqq 
	&\frac{1}{16}\sum_{i<j}\Big(\frac{1}{\sin^2\frac{\pi}{2N}(i-j)}(1-P_{ij})\\
	&+\frac{1}{\cos^2\frac{\pi}{2N}(i-j)}(1-P_{ij}\sigma_i^x\sigma_j^x)\Big)\, , 
	\end{aligned} 
\end{equation}
which upon an irrelevant constant shift and rescaling coincides with equation (2.1) in \cite{basu-mallickNovelClassTranslationally2020}. 

We can rewrite the hamiltonian \eqref{eq:HBM} further as follows: 
\begin{align}\label{eq:HBMrewrite}
&	H_{\textsc{bmfgl}}=
	%&=\frac{1}{4}\sum_{i<j}\left[\frac{1}{\sin^2\tfrac{\pi}{2N}(i-j)}(1-P_{ij})+\frac{1}{\cos^2\tfrac{\pi}{2N}(i-j)}(1-P_{ij}\sigma_i^x\sigma_j^x)\right]\nonumber\\	
\nonumber	\\
	&=\sum_{i<j}\frac{1}{4\sin^2\tfrac{\pi}{N}(i-j)}\big(\cos^2\big(\tfrac{\pi}{2N}(i-j)\big)(1-P_{ij}) \nonumber \\
	&+\sin^2\big(\tfrac{\pi}{2N}(i-j)\big)(1-P_{ij}\sigma_i^x\sigma_j^x)\big)\nonumber\\
%	&=\sum_{i<j}\frac{1}{\sin^2\frac{\pi}{N}(i-j)}\left(1-P_{ij}-\sin^2\frac{\pi}{2N}(i-j)P_{ij}(\sigma_i^x\sigma_j^x-1)\right)\nonumber\\
	&=\sum_{i<j}\frac{1}{4\sin^2\frac{\pi}{N}(i-j)}\big(1-P_{ij}\nonumber\\
	&+\sin^2\big(\tfrac{\pi}{2N}(i-j)\big)(\sigma_i^y\sigma_j^y+\sigma_i^z\sigma_j^z)\big)\\
	&= \sum_{i<j}\frac{1}{4\sin^2\tfrac{\pi}{N}(i-j)}\Big((1-P_{ij}) \nonumber \\
	&+\frac{1-\cos\frac{\pi}{N}(i-j)}{2}(\sigma_i^y\sigma_j^y+\sigma_i^z\sigma_j^z)\Big)  \nonumber \\
	&=-\frac{1}{2}\sum_{i<j}\frac{1}{4\sin^2\frac{\pi}{N}(i-j)}\Big(\sigma_i^x\sigma_j^x-1  \nonumber \\
	&+\cos\frac{\pi}{N}(i-j)(\sigma_i^y\sigma_j^y+\sigma_i^z\sigma_j^z)\Big)\nonumber \, , 
\end{align}
using $\sin2x=2\sin x\cos x$ and $P_{ij}(1-\sigma_i^x\sigma_j^x)=\sigma_i^y\sigma_j^y+\sigma_i^z\sigma_j^z$. Also performing the (Hadamard) change of basis defined by the matrix
\begin{equation}
	\label{eq:hadamard}
	U_H=\bigotimes_{i=1}^N\frac{1}{\sqrt{2}}
	\begin{pmatrix}
		1 & -1\\
		1 & 1
	\end{pmatrix}\, , 
\end{equation}
which amounts to swapping $\sigma_i^x \to -\sigma_i^z$ and $\sigma_i^z \to \sigma_i^x$, we obtain the hamiltonian
\begin{equation}
	\begin{aligned}
	\label{eq:HFK12}
		&U_H H_{\text{BMFGL}}U_H^{-1} =\\ &=-\sum_{i<j}\frac{1}{4\sin^2\frac{\pi}{N}(i-j)}\times  \\
		&\Big(\cos\frac{\pi}{N}(i-j)(\sigma_i^+\sigma_j^-+\sigma_i^-\sigma_j^+)+\frac{\sigma_i^z\sigma_j^z-1}{2}\Big)\, ,
		\end{aligned}
\end{equation}
which coincides with the FK hamiltonian \eqref{eq:HFKdef} for $\varphi = 1/2$. It also equals the trigonometric hamiltonian obtained by Sechin and Zotov in \cite{sechin2018r} by considering \textit{R}-matrix valued Lax pairs, and fits in the landscape of integrable spin chains generated by taking all possible limits of the elliptic Matushko--Zotov chain \cite{MZ_23a}. We hence conclude that
%, contrary to a hypothesis in \cite{basu-mallickNovelClassTranslationally2020}, 
$H_\textsc{fk}(1/2)$ is equivalent to both these other hamiltonians \footnote{For completeness, let us mention that mirroring in the axis separating the two chain halves -- as opposed to the antipodal mirroring in Fig.~\ref{fg:three_chains}-- yields yet another (but truly inequivalent) long-range integrable spin chain, namely the $BC_N$-invariant chain introduced in \cite{bernardExactSolutionLongRange1995}.}. Following \textsection\ref{sec:twisted_models}, the (unnormalised) translation operator that commutes with $H_{\text{BMFGL}}$ takes the form 
\begin{equation}\label{eq:translation12}
	U_H T(1/2) U_H^{-1} = \sigma_N^x T(0) = T(0)  \sigma_1^x\, ,
\end{equation}
indeed coinciding with the translation operator from \cite[p.7]{basu-mallickNovelClassTranslationally2020}. 

\begin{figure*}
	\centering
	\begin{tikzpicture}[scale=0.45]
    % Draw y-axis on the left
    \draw[<-, >={Stealth[length=6pt, width=6pt]}] (-1,-3.6) -- (-1,3.6);
    \foreach \y in {-3,...,3} {
    \draw (-1.1,\y) -- (-0.9,\y);
    \node[left] at (-1.4,\y) {\y};
    }
    \node at (-1,-4.3) {$S^z$};
    
    \draw[->, >={Stealth[length=6pt, width=6pt]}] (-1,3.6) -- (32,3.6);
    \node at (33.,3.6) {$4E_{s\textsc{hs}}$};
  
  % Lines for each representation
  \foreach \i/\k in {
    0/2,
    1/1, 2/1, 3/1, 4/1, 5/1,
    6/0, 7/0, 8/0, 9/0, 10/0, 11/0, 12/0, 13/0, 14/0, 15/0,
    16/-1, 17/-1, 18/-1, 19/-1, 20/-1, 21/-1, 22/-1, 23/-1, 24/-1, 25/-1,
    26/-2, 27/-2, 28/-2, 29/-2, 30/-2,
    31/-3
  } {
    \draw (\i,\k) -- (\i,\k+1);
    \fill (\i,\k+1) circle (5pt);
    \fill (\i,\k) circle (5pt);
  }
  
  % Motifs at the bottom
  \foreach \i/\m in {
    0/{(0)},
    1/{(1)}, 2/{(5)}, 3/{(2)}, 4/{(4)}, 5/{(3)},
    6/{(1,5)}, 7/{(4,5)}, 8/{(2,5)}, 9/{(1,4)}, 10/{(1,2)},
    11/{(1,3)}, 12/{(3,5)}, 13/{(2,4)}, 14/{(3,4)}, 15/{(2,3)},
    16/{(1,4,5)}, 17/{(1,2,5)}, 18/{(1,3,5)}, 19/{(2,4,5)}, 20/{(1,2,4)},
    21/{(3,4,5)}, 22/{(2,3,5)}, 23/{(1,3,4)}, 24/{(1,2,3)}, 25/{(2,3,4)},
    26/{(1,2,4,5)}, 27/{(1,3,4,5)}, 28/{(1,2,3,5)}, 29/{(2,3,4,5)}, 30/{(1,2,3,4)},
    31/{(1,2,3,4,5)}
  }{
    \node[rotate=70, anchor=east, font=\tiny] at (\i+0.2,-3.2) {\m};
  }
  
  % Energy labels at height 4
  \foreach \i/\e in {
    0/0,
    1/5, 2/5, 3/8, 4/8, 5/9,
    6/10, 7/13, 8/13, 9/13, 10/13, 11/14, 12/14, 13/16, 14/17, 15/17,
    16/18, 17/18, 18/19, 19/21, 20/21, 21/22, 22/22, 23/22, 24/22, 25/25,
    26/26, 27/27, 28/27, 29/30, 30/30,
    31/35
  } {
    \node[font=\small] at (\i,4) {\e};
  	}
	\end{tikzpicture}
	\caption{Schematic picture of the Hilbert space for the supersymmetric chain for length $N=6$. It consists of $\mathfrak{gl}(1|1)$-doublets only, which are labelled uniquely by a supermotif as indicated on the bottom and four times their energy on the top. For easy comparison we have organised the spectrum using its decomposition into $S^z$-eigenspaces rather than the $\mathfrak{gl}(1|1)$-native fermionic number operator $N_G = N/2-S^z$ from \eqref{eq:ferm_gens}. Compare with the bottom half of Fig.~\ref{fg:FKvsHS_chain_picture}, which uses a different method to organise the same spectrum.}
	\label{fg:susy_chain_picture}
\end{figure*}
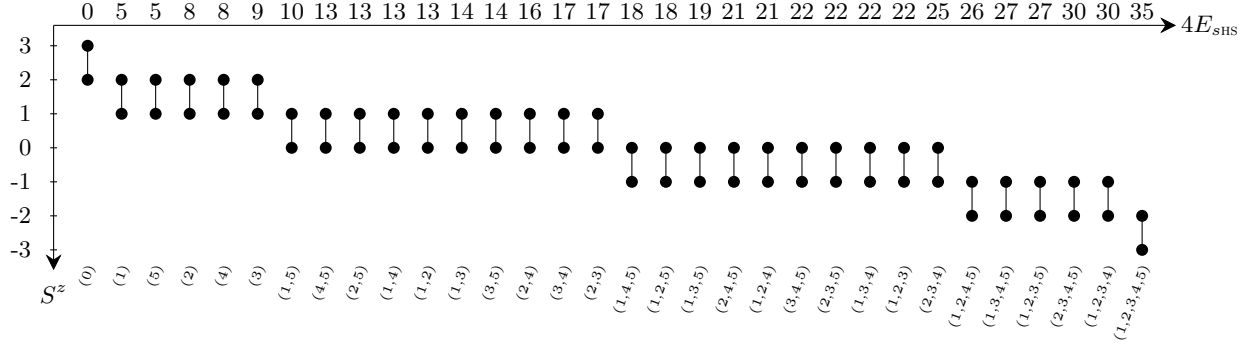

As proven in \cite{basu-mallickNovelClassTranslationally2020} by examining the partition function of $H_{\text{BMFGL}}$, we can add a third chain to the list of equivalent chains: the supersymmetric $\mathfrak{gl}(1|1)$ Haldane--Shastry chain \cite{Hal_94} defined by the hamiltonian
\begin{equation}
	\label{eq:susy_HS}
	H_{\text{s}\textsc{hs}} = \frac{1}{2}\sum_{i<j}^N \frac{1-\mathbb{P}_{ij}}{4 \sin^2 \!\tfrac{\pi}{N}(i-j)}\, , 
\end{equation}
where $\mathbb{P}_{ij}$ is the supersymmetric spin-exchange operator taking care of the extra minus signs upon exchanging fermions, and can be realised as
\begin{equation}
	\mathbb{P}_{ij} = (f_i - f_j)(f_i^\dagger-f_j^\dagger)-1\, . 
\end{equation}
in terms of the Jordan-Wigner transformed Pauli matrices \footnote{\label{fn:jordan_wigner} As an aside, we note that per the Jordan-Wigner transformation one can interpret the $f_j$ and $f_j^\dagger$ as 'dynamical' spin chain operators, in the sense of dynamical quantum groups \cite{felder1995conformal}. Indeed, the products over $\sigma^z$s measure the spins on all sites to the left of the acting operator, and can be written as a shift in the exponent $\exp( \I \pi \sum_{k=1}^{j-1} \sigma_k^z)$.}
\begin{equation}
	f_j = \left(\prod_{k=1}^{j-1} \sigma^z_k \right) \sigma_j^+\, , \quad 	f_j^\dagger = \left(\prod_{k=1}^{j-1} \sigma^z_k \right) \sigma_j^-\, . 
\end{equation}
We give a full elementary description of this model in Appendix \ref{app:susy_HS}, and list its pertinent properties here: the chain \eqref{eq:susy_HS} has a global $\mathfrak{gl}(1|1)$-symmetry generated by
\begin{equation}
	\label{eq:ferm_gens}
	\begin{aligned}
	N_G &= \sum_{i=1}^N f_i^\dagger f_i\, , \quad E_G = 1 \, , \quad F_G= \sum_{i=1}^N f_i \, , \\
	 F^\dagger_G &=  \sum_{i=1}^N f_i^\dagger\, ,
	\end{aligned} 
\end{equation}
which ensures its spectrum is very simple. It can be constructed from highest-weight states characterised by a $\mathfrak{gl}(1|1)$- (or super-)motif \cite{Hal_94}, a set of integers $\nu_i \in \{1,\ldots, N-1\}$ satisfying $\nu_i <\nu_{i+1}$ (compare with \eqref{eq:motif_relation}), encoding the eigenvalues
\begin{equation}
	\label{eq:susy_eigenvalues}
	\begin{aligned}
	E_{\text{s}\textsc{hs}} &= \sum_{i} \frac{\epsilon_0(\nu_i)}{2}\, , \quad P = \frac{2 \pi}{N} \sum_{i} \nu_i\, , \\
	\lambda^z &= \sum_{i} (N/2-\nu_i)\, , 
	\end{aligned}
\end{equation}
where we stress that the energy is a sum of single-particle energies parametrised by \emph{half} the dispersion of the ($\mathfrak{gl}(2|0)$-) HS-chain. Also, $\lambda^z$ is the eigenvalue of the extra commuting charge
\begin{equation}
	Q^z_{\text{s}\textsc{hs}} = \frac{1}{2} \sum_{i< j} \cot \frac{\pi}{N} (i-j) \left( f_i^\dagger f_j - f_j^\dagger f_i \right)\, , 
\end{equation} 
extending the full symmetry to $Y(\mathfrak{gl}(1|1)$. From the highest-weight state one can produce one descendant by acting with $F_G^\dagger$, but since it squares to zero no further states can be reached: the spectrum consists solely of $\mathfrak{gl}(1|1)$-doublets (see Fig.~\ref{fg:susy_chain_picture}). 

This fact is not only highly non-trivial, but also very interesting. Indeed, the analysis in \cite{basu-mallickNovelClassTranslationally2020} of the partition function shows that for these two chains the eigenvalues and their multiplicities coincide \footnote{The analysis in \cite{basu-mallickNovelClassTranslationally2020} treats the general rank case; setting $m=2$ one retrieves from equation (4.10) therein that the partition functions of $H_{\text{BMFGL}}$ and  $H_{\text{sHS}}$ coincide.}. This implies there exists a change of basis that makes the equivalence explicit. Comparisons for small lengths indicate that such a transformation does not preserve magnon numbers, i.e. it maps some eigenvectors to eigenvectors with a different magnon number. This makes a brute force approach to finding such a basis change impractical. 

\section{Supermotifs in terms of motifs}
We have established that $H_{\text{FK}}(1/2)$ is related to $H_{\textsc{bmfgl}}$ by a simple basis transformation, and know from \cite{basu-mallickNovelClassTranslationally2020} that the latter is equivalent to $H_{s\textsc{HS}}$. Since the latter has a $\mathfrak{gl}(1|1)$-symmetry, the same must be true for the other chains. Of course, this means that the spectra of these chains are organised in $\mathfrak{gl}(1|1)$-doublets (see Fig.~\ref{fg:susy_chain_picture}), providing a very different way of decomposing the spectrum compared to that in Fig.~\ref{fg:FKvsHS_chain_picture}. The doublets come with a natural labelling using supermotifs, but whether one can interpret these supermotifs in more physical terms remains to be understood. More concretely, the spinons of the $H_{s\textsc{HS}}$ are \emph{free} fermions (which should not be confused with the fermions occupying a fixed site and which interact as described by the hamiltonian), and hence these other chains should similarly allow for a description in terms of excitations that only interact through their (fermionic) exchange statistics. This mimics the case of the HS chain, which in the same way is a model of free semions. Identifying the fermionic excitations in $H_{\text{FK}}(1/2)$ would not only explain its hidden simplicity, but also constitute progress towards understanding the (anyonic) excitations of the FK-chains for other values of $\varphi$. 

A step towards understanding this is to explicitly relate multiplets from the different chains. Given a $\mathfrak{gl}(2|0)$-motif $\mu$ labelling a state of $H_{\text{FK}}(1/2)$ through the mechanism explained in \textsection\ref{sec:twisted_and_untwisted}, we can find the associated $\mathfrak{gl}(1|1)$-motif $\nu$ labelling a $H_{s\textsc{HS}}$-state with the same energy by `thickening' the motif
\begin{equation}
	\label{eq:unsqueezing}
	\begin{aligned}
	&(\mu_1,\mu_2,\ldots, \mu_M) \mapsto \nu_\mu =\\
	 &=(\mu_1, \mu_1+1,\mu_2,\mu_2+1,\ldots \mu_M,\mu_M+1) \text{ mod } N\, .
	\end{aligned}
\end{equation}
This operation also appeared in \cite{jiangNormsOverlapsYangian2026}, but not in relation to supermotifs. The motif condition \eqref{eq:motif_relation} ensures that the right-hand side does not have repeated entries, but may include a (single) zero due to $N \equiv 0 \text{ mod }N$. It thus takes the form of a supermotif, where supermotifs $\nu' = \{0\} \cup \nu$ containing a zero denote the unique descendant of the highest weight state with supermotif $\nu$. A geometric, but heuristic, way to understand `thickening' is by looking again at the wrapping procedure in \textsection\ref{sec:twisted_models} that led to the half-twisted hamiltonian $H_{\text{FK}}(1/2)$: it is obtained by wrapping twice around the chain with a half-turn after the first time around. For a state with motif $\mu$ this yields two entries for each $\mu_i$, first at $\mu_i$ and then at $\mu_i+\varphi = \mu_i+1/2 \rightsquigarrow \mu_i+1$, with the first shift due to the twist and the rounding occurring because the $\mu_i$ cannot be coinciding and must be integer.

The dispersion relation admits a similar interpretation, as it satisfies
\begin{equation}
	\varepsilon_{1/2}(\mu) =  \frac{\varepsilon_{0}(\mu)  + \varepsilon_{0}(\mu+1)}{2}\, , 
\end{equation}
indicating that a single excitation of the twisted chain can be split into a (weighted) sum of two excitations of the untwisted one. Inspecting \eqref{eq:susy_eigenvalues} we see moreover that we can reinterpret the right-hand side as an (now unweighted) sum of single-particle energies of the $\mathfrak{gl}(1|1)$-chain. It is not difficult to see that this extends to multi-particle states, and the associated energies of two  motifs related by \eqref{eq:unsqueezing} coincide as well: 
\begin{equation}
	E_\textsc{fk} (\mu) = 	E_{\text{s}\textsc{hs}}(\nu_\mu)\, . 
\end{equation}
Thus the map \eqref{eq:unsqueezing} allows us to relate part of the spectrum of $H_{\text{FK}}(1/2)$ to the representation theory of the free fermion model $H_{s\textsc{HS}}$. Note that it inherently relates states with different magnon numbers (equalling the number of components of the motif), and even sends Yangian highest weight states to descendants in the fermionic chain. Moreover, it is clear that this map also relates two different notions of momentum: for small system size one can check that the translation operators $\bar{T}(1/2)$ and its supersymmetric counterpart defined in \eqref{eq:susy_translation} are not related by the change of basis, indicating there should be a second translation operator for $H_{\text{FK}}(1/2)$ which measures the momentum of the so far unknown excitations. 
\begin{comment}
\begin{figure}[h!]
\begin{tikzpicture}[scale=0.4]
%Sz axis
\draw[->, >={Stealth[length=6pt, width=6pt]}] (-1,-3.6) -- (-1,+3.6);
\pgfmathsetmacro{\ymin}{-3}
\pgfmathsetmacro{\ymax}{3}
\foreach \y in {\ymin,...,\ymax} {
  \draw (-1.1,\y) -- (-0.9,\y);
  \pgfmathsetmacro{\ylabel}{\y}
\node[left] at (-1.4,\y) {\pgfmathprintnumber{\ylabel}};
}
\node at (-1,4.3) {$S^z$};
%Quadruplet
\draw (1,-4/2) -- (1,4/2);
\fill (2,-4/2) circle (5pt) node[right] {$\frac{\pi}{6}$};
\fill (2,4/2) circle (5pt) node[right] {$\frac{\pi}{6}$};
\draw (2,-4/2) -- (2,4/2);
\fill (1,-4/2) circle (5pt) node[left] {$\frac{5\pi}{6}$};
\fill (1,4/2) circle (5pt) node[left] {$\frac{5\pi}{6}$};
\draw[dashed] (1,4/2) -- (2,4/2);
\draw[dashed] (1,-4/2) -- (2,-4/2);
\node[align=center] at (1.5,-4) {5};  
\end{tikzpicture}
\end{figure}
\end{comment}

\section{Outlook}
We have analysed the spectrum of Fukui and Kawakami's twisted long-range spin chains utilising a connection to the periodic Haldane--Shastry chain. Our results indicate that these models harbour a lot of hidden structure and suggest many possible paths for future research. 

Understanding the precise nature of the extra descendants in the two magnon sector would be very interesting, e.g by constructing a set of (generalised) lowering operators which deform the Yangian lowering operators $S^-$ and $Q^-$ of the HS chain. Further mapping out how these generalise to higher magnon sectors forms an important stepping stone towards uncovering the algebraic origins of the FK chain's solvability. The case of $\varphi=1/2$ deserves special attention, being equivalent to three previously studied chains, as it has a hidden $\Z_2$- or supersymmetry. The underlying mechanism could be similar to well-known cases of supersymmetry in short-range chains \cite{yangNonlocalSpacetimeSupersymmetry2004,hagendorfEightvertexModelLattice2012}. 

The chains studied here are more complicated than the HS chain, but considerably simpler than other integrable anisotropic long-range chains such as the $q$-deformed HS chain \cite{uglovTrigonometricCounterpartHaldane1995,Lam_18} or its six-vertex counterpart obtained by a trigonometric limit from the chain defined in \cite{MZ_23a}, see \cite{klabbers2024landscapes}. Yet, the FK chains are not isotropic, and could therefore be candidates to model realistic behaviour, for example in Rydberg atoms. Moreover, they are a natural testing ground to further investigate the interplay between ballistic transport and long-range interactions \cite{mierzejewskiQuasiballisticTransportLongrange2023,anandRobustnessKardarParisiZhanglikeTransport2026}. 

Since the FK chain at $\varphi =1/2$ is equivalent to a model of free fermions, and the HS chain (corresponding to $\varphi=0$) is a model of free semions, it is natural to conjecture that for other $\varphi$ the spinons of the FK chain are (free) anyons with their exchange parameter depending on $\varphi$. This would make the FK chains the first family of integrable spin-$1/2$ chains to model anyons of varying exchange parameter and unlock the potential to study their properties analytically. 

A further step would be to consider whether there are natural deformations of the interaction potential to allow for variation in its range without breaking integrability. Such a model exists, and can be obtained from the $q$-deformed Inozemtsev spin chain \cite{klabbersDeformedInozemtsevSpin2024} by taking a well-tuned limit, and features an elliptic interaction potential. As will expanded upon elsewhere, the short-range limit of that elliptic twisted chain is precisely the twisted Heisenberg \textsc{xxx} spin chain discussed in \textsection\ref{sec:twisted_models}, further underlining that the FK chains are \emph{twistings} of the HS chain \cite{Thomas}. 

Another direction would be to investigate whether the identity \eqref{eq:HFKHHSrel} has a $q$-deformed analogue, featuring a suitably twisted version of the $q$-deformed Haldane--Shastry chain \cite{Ugl_95u,Lam_18}. Such an identity provides further connections to explore the representation theory of quantum groups, and could help explain the structure appearing at roots of unity, such as at $q=\I$ \cite{benmoussaSolvableNonunitaryFermionic2025}. 

In addition to providing simple integrable models with which to study the various physical phenomena introduced above, the FK chains are very interesting from a mathematical perspective as well. The hidden structure on which their solvability relies should deform the Yangian representation theory of the HS chain \cite{bernard1993yang}, with natural connections to orthogonal (Jack) polynomials and combinatorics. Moreover, just like Sklyanin's boundary Yang-Baxter equation or Drinfel'd twistings for short-range chains, the putative algorithm that generates the necessary algebraic structure can be used to produce many other integrable models. Finally, the construction of the twisted spectrum is likely to prove a useful tool to understand the untwisted spectrum, cf. \cite{jiangNormsOverlapsYangian2026}. 
\section*{Acknowledgements}
We thank Andrii Liashyk for discussions and sharing his joint result with Ivan Sechin of the identity \eqref{eq:HFKHHSrel} on the $\varphi=1/2$ case, as well as the organisers of RAQIS '24 in Annecy (France) where these discussions took place. We thank the authors of \cite{basu-mallickNovelClassTranslationally2020} -- Bireswar Basu-Mallick, Federico Finkel, and Artemio González-López -- for correspondence. We further thank Sibylle Driezen, Julio Cabello Gil, Ana Retore, and Didina Serban for their interest and discussions, and Jules Lamers and Thomas Scheutz for discussions and comments on the manuscript. A.L. thanks Humboldt Universität zu Berlin for their hospitality during part of the work. R.K.’s research is partially funded
by the Deutsche Forschungsgemeinschaft (DFG, German Research Foundation) – Projektnummer 417533893/GRK2575 “Rethinking Quantum Field Theory”.
\appendix
\section{Wrapping identity}
\label{app:trig_identity}
The fundamental wrapping identities for the Haldane--Shastry chain are formulae for periodisations of rational functions. For example, the double pole $1/z^2$ gets wrapped into an $N$-periodic function as 
\begin{equation}
	\sum_{k\in \Z} \frac{1}{(z+k N)^2} = \frac{(\pi/N)^2}{\sin^2 \tfrac{\pi}{N} z}\, , 
\end{equation}
the interaction potential of the periodic chain. Periodising the simple pole $1/z$ is famously a little tricky, as the convergence of 
\begin{equation}
	\sum_{k\in \Z} \frac{1}{z+k}
\end{equation}
is conditional; it requires a prescription in which order this sum is to be computed, such as 
\begin{equation}
	\sum_{k\in \Z} \frac{1}{z+k} = \frac{1}{z} + \sum_{k\geq 1} \Bigg(\frac{1}{z+k} + \frac{1}{z-k} \Bigg)\, . 
\end{equation}
The sum on the right-hand side now converges uniformly and absolutely on compacts away from its poles. To compute it one can perform the contour integral of the function $f(w) = \cot \pi w \frac{1}{z+w}$ on a circular contour of increasing radius. On the one hand $f$ is well balanced on this contour, and the integral vanishes. On the other hand, by deforming the contour one sees that the integral also equals an infinite sum of residues around its poles. This yields
\begin{equation}
	\label{eq:periodic_cot}
 \frac{1}{z} + \sum_{k\geq 1} \Bigg(\frac{1}{z+k} + \frac{1}{z-k} \Bigg) = \pi \cot \pi z\, . 
\end{equation}
Using this identity we can now derive a finite periodisation of the cotangent, taking it from a $q$-periodic function to a $1$-periodic one: 
\begin{equation}
	\label{eq:cot_periodisation}
	\begin{aligned}
	&\frac{1}{q}\sum_{m=0}^{q-1} \cot \tfrac{\pi}{q} (z+m)= \frac{1}{\pi}\sum_{m=0}^{q-1} \Bigg( \frac{1}{z+m} + \\
	&\sum_{k\geq 1} \Bigg(\frac{1}{z+m+k q} + \frac{1}{z+m-kq} \Bigg)\Bigg) \\
	&= \frac{1}{\pi}\Bigg( \frac{1}{z} + \sum_{l\geq 1} \Bigg(\frac{1}{z+l} + \frac{1}{z-l} \Bigg)\Bigg) =\cot \pi z\, , 
	\end{aligned}
\end{equation}
making use of the fact that the $m\pm kq$ combine to cover $\Z$ precisely once to allow for a relabelling. The identity \eqref{eq:cottwisted} (repeated here)
\begin{displaymath}
			\frac{1}{q}\sum_{m=0}^{q-1} e^{2\pi \I \frac{p}{q}(z+m)}\cot\frac{\pi}{q}(z+m) = \frac{e^{\pi\I z}}{\sin\pi z}\, ,
\end{displaymath}
which we aim to prove here, is a twisting of the identity \eqref{eq:cot_periodisation}. 

Repeating the first steps, we can rewrite the finite sum as 
\begin{align}\label{eq:cottwistedrewrite}
	&\frac{1}{q}\sum_{m=0}^{q-1} e^{2\pi \I \frac{p}{q}(z+m)}\cot\frac{\pi}{q}(z+m) \\
	&
	= \frac{e^{2\pi \I \frac{p}{q}z}}{\pi} \sum_{k\in \mathbb{Z}} \sum_{m=0}^{q-1} \frac{e^{2\pi \I \frac{p}{q}m}}{z+m+qk} \\
	&= \frac{e^{2\pi \I \frac{p}{q}z}}{\pi}\sum_{k\in \mathbb{Z}} \sum_{m=0}^{q-1} \frac{e^{2\pi \I \frac{p}{q}(m+qk)}}{z+m+qk} \nonumber \\
	&= \frac{e^{2\pi \I \frac{p}{q}z}}{\pi}\sum_{n\in \mathbb{Z}}  \frac{e^{2\pi \I \frac{p}{q}n}}{z+n}\, ,
\end{align}
where we interpret the infinite sum as pairing the summands at $k$ and $-k$, as before. 

Introducing the function
\begin{equation}
	f(w)=\frac{1}{e^{2\pi\I w}-1}\frac{e^{2\pi \I \frac{p}{q}w}}{z+w}
\end{equation}
for some $z\in \C\setminus\Z$, 
we would like to perform the exact same contour integral as before:
\begin{equation}
	\oint_{C} \frac{dw}{2\pi \I} f(w)\, ,
\end{equation}
Since the function $f(w)$ is periodic along the real axis, we choose a square contour where the two vertical sides are separated by an integer number of the period of the function. This ensures that the contributions from these vertical sides cancel each other out. For the two horizontal lines, $f(w)$ decays exponentially:
\begin{equation}
\frac{e^{2\pi\I\frac{p}{q}w}}{e^{2\pi\I w}-1} \propto \frac{e^{-\frac{p}{q}\Im w}}{e^{- \Im w}-1} \simeq
\left\{
    \begin{array}{ll}
        e^{-\frac{p}{q}\Im w} \rightarrow 0 \\
      e^{(1-\frac{p}{q})\Im w} \rightarrow 0 
    \end{array}
\right.
\end{equation}
with the upper line indicating the behaviour for $\mathrm{Im}\, w > 0$ and the lower line that for $\mathrm{Im} \, w <0$ 

Hence the contour integral vanishes, but on the other hand it must equal the sum of its residues: 
\begin{equation}
-\pi e^{\pi\I z} \frac{e^{-2\pi \I \frac{p}{q}z}}{\sin \pi z}+\sum_{n\in\mathbb{Z}} \frac{e^{2\pi \I \frac{p}{q}n}}{z+n} =0\, , 
\end{equation}
with the first term constituting the residue at $w= -z$. Rearranging and substituting into \eqref{eq:cottwistedrewrite}:
\begin{align}
\frac{1}{q}\sum_{m=0}^{q-1} & e^{2\pi \I \frac{p}{q}(z+m)}\cot\frac{\pi}{q}(z+m) \nonumber 
\\
&= \frac{1}{\pi} e^{2\pi \I \frac{p}{q}z} \sum_{n\in \mathbb{Z}}  \frac{e^{2\pi \I \frac{p}{q}n}}{z+n}\nonumber \\
&= e^{2\pi \I \frac{p}{q}z} e^{\pi\I z} \frac{e^{-2\pi \I \frac{p}{q}z}}{\sin \pi z} = \frac{e^{\pi\I z}}{\sin\pi z}
\end{align}
we get the identity \eqref{eq:cottwisted} we were after.

Taking a derivative with respect to $z$ leads the other identity used in obtaining \eqref{eq:HFK+QFKbig}. The left hand side of \eqref{eq:cottwistedrewrite} gives:
\begin{align}
\frac{d}{dz}&\left[\frac{1}{q}\sum_{m=0}^{q-1} e^{2\pi \I \frac{p}{q}(z+m)}\cot\frac{\pi}{q}(z+m)\right] \nonumber \\
&=
\frac{1}{q}2\pi\I\frac{p}{q}\sum_{m=0}^{q-1} e^{2\pi \I \frac{p}{q}(z+m)}\cot\frac{\pi}{q}(z+m) \nonumber  \\
&- \frac{\pi}{q^2}\sum_{m=0}^{q-1} \frac{e^{2\pi \I \frac{p}{q}(z+m)}}{\sin^2\frac{\pi}{q}(z+m)}\nonumber \\
&= 2\pi\I\frac{p}{q}\frac{e^{\pi\I z}}{\sin\pi z} - \frac{\pi}{q^2}\sum_{m=0}^{q-1} \frac{e^{2\pi \I \frac{p}{q}(z+m)}}{\sin^2\frac{\pi}{q}(z+m)}
\end{align}
and the right hand side:
\begin{equation}
	\begin{aligned}
\frac{d}{dz}\left[\frac{e^{\pi\I z}}{\sin\pi z}\right]&=\frac{\pi\I e^{\pi\I z}\sin\pi z-e^{\pi\I z}\pi\cos\pi z}{\sin^2\pi z}\\
&=-\pi\frac{1}{\sin^2\pi z}
\end{aligned}
\end{equation}
hence we obtain
\begin{equation}
\frac{1}{q^2}\sum_{m=0}^{q-1} \frac{e^{2\pi \I \frac{p}{q}(z+m)}}{\sin^2\frac{\pi}{q}(z+m)} = \frac{1}{\sin^2\pi z} + 2\I\frac{p}{q}\frac{e^{\pi\I z}}{\sin\pi z}\, . 
\end{equation}

\section{Supersymmetric Haldane--Shastry chain}
\label{app:susy_HS}
The hamiltonian \eqref{eq:susy_HS} models long-range interactions between spinless fermions. The supersymmetric spin-exchange operator $\mathbb{P}_{ij}$ can be defined as follows: for neighbouring sites $i$ and $i+1$ we set
\begin{equation}
	\mathbb{P}_{i,i+1} \coloneqq 1 \otimes \ldots \otimes 
	\begin{pmatrix}
		1 & \gz & \gz & \gz \\
		\gz & 0 & 1 & \gz \\
		\gz & 1 & 0 & \gz \\
		\gz & \gz & \gz & \scalebox{0.7}{$-$}1
	\end{pmatrix} \otimes \ldots \otimes 1\, , 
\end{equation}
which incorporates a minus sign for the exchange of two fermions. These operators satisfy the defining relations of the symmetric group $S_N$, 
\begin{equation}
	\label{eq:sym_group_relns}
	\begin{aligned}
	\mathbb{P}_{i,i+1}^2 &= 1\, , \\
	 \mathbb{P}_{i,i+1} \mathbb{P}_{i+1,i+2}\mathbb{P}_{i,i+1} &= \mathbb{P}_{i+1,i+2}\mathbb{P}_{i,i+1}\mathbb{P}_{i+1,i+2}\, , 
	 \end{aligned}
\end{equation}
thus they generate an $S_N$-representation. To find an expression for $\mathbb{P}_{ij}$ for any $i$ and $j$, we first decompose the associated transposition $(ij) \in S_N$ into nearest-neighbour (or 'elementary') transpositions. Despite there being multiple ways of performing this decomposition, all choices are equivalent by virtue of the braid relation in \eqref{eq:sym_group_relns}. For concreteness, for $i<j$ we pick the reduced decomposition $(ij) = (j-1\,  j)\cdots (i+1\, i+2)(i \,i+1) (i+1 \, i+2) \cdots (j-1 \, j)$ and define
\begin{equation}
	\mathbb{P}_{ij} \coloneqq \ordprod_{k=i+1}^{j-1} \mathbb{P}_{k\, k+1} \cdot \ordprodopp_{k=i+1}^{j} \mathbb{P}_{k-1\, k}\, , 
\end{equation}
with the harpoon indicating the ordering of the factors. Introducing the diagram
\begin{equation} \label{eq:deformed_permutation_diagram}
	\mathbb{P}_{i,i+1} = \,
	\tikz[baseline={([yshift=-.5*11pt*0.3]current bounding box.center)},xscale=.6,yscale=0.3,font=\footnotesize]{
		\draw[->] (0,0) node[below]{$1$} -- (0,3);
		\foreach \x in {-1,...,1} \draw (.75+.2*\x,1.5);	
		\draw[->] (1.5,0) -- (1.5,3);
		\foreach \x in {1,2,3} 
		\draw (1.1+.2*\x,-0.9) node{$\cdot\mathstrut$}
		;
		\draw[rounded corners=2pt,->] (3.5,0) node[below]{$i+1$} -- (3.5,1) -- (2.5,2) -- (2.5,3) ;
		\draw[rounded corners=2pt,->] (2.5,0) node[below]{$i$} -- (2.5,1) -- (3.5,2) -- (3.5,3);
		\draw[->] (4.5,0) -- (4.5,3);
		\foreach \x in {1,2,3} 
		\draw (4.15+.2*\x,-0.9) node{$\cdot\mathstrut$}
		;
		\foreach \x in {-1,...,1} \draw (5.25+.2*\x,1.5);	
		\draw[->] (6,0) node[below]{$N$} -- (6,3);
	}  \, ,
\end{equation}
with the line subscripts labelling the site from which the line departs, we can depict $\mathbb{P}_{ij}$ as follows: 
\begin{equation}
	\mathbb{P}_{ij} = \,\,
	\tikz[baseline={([yshift=5pt]current bounding box.center)},xscale=0.4,yscale=0.2,font=\footnotesize]{
		\draw[->] (0,0) node[below]{$1$} -- (0,9);
		\draw[->] (1.5,0) node[below]{$i$} -- (1.5,3.75) .. controls (1.5,4.1) and (1.75,4.25) .. (2,4.5) -- (5,7.5).. controls (5.25,7.75) and (5.5,7.9) ..(5.5,8.25) -- ((5.5,9);
		\draw[->] (2.5,0) -- (2.5,2.75) .. controls (2.5,3.5) and (3.5,3.5) .. (3.5,4.25) -- (3.5,4.75) .. controls (3.5,5.5) and (2.5,5.5) .. (2.5,6.25) -- (2.5,9);
		\draw[->] (3.5,0) -- (3.5,1.75) .. controls (3.5,2.5) and (4.5,2.5) .. (4.5,3.25) -- (4.5,5.75) .. controls (4.5,6.5) and (3.5,6.5) .. (3.5,7.25) -- (3.5,9);
		\draw[->] (4.5,0) -- (4.5,0.75) .. controls (4.5,1.5) and (5.5,1.5) .. (5.5,2.25) -- (5.5,6.75) .. controls (5.5,7.5) and (4.5,7.5) .. (4.5,8.25) -- (4.5,9);
		\draw[->] (5.5,0) node[below]{$j$} -- (5.5,0.75) .. controls (5.5,1.1) and (5.25,1.25) .. (5,1.5) -- (2,4.5) .. controls (1.75,4.75) and (1.5,4.9) .. (1.5,5.25) -- (1.5,9);
		\draw[->] (7,0) node[below]{$N$} -- (7,9);
		\node at (0.75,4.5) {$\cdots$};
		\node at (6.25,4.5) {$\cdots$};
	}
\end{equation}
In particular, the (left)-translation operator can be built out of the nearest-neighbour $\mathbb{P}_{i,i+1}$ by the product 
\begin{equation}
	\label{eq:susy_translation}
	\mathbb{G} = \,\,
	\tikz[baseline={([yshift=-.5*11pt*0.2]current bounding box.center)},xscale=0.4,yscale=0.2,font=\footnotesize]{
		\draw[rounded corners=2pt,->] (0,0) node[below]{$1$} -- (0,1) -- (5,6) -- (5,7.5);
		\draw[rounded corners=2pt,->] (1,0) node[below]{$2$} -- (1,1) -- (0,2) -- (0,7.5);
		\foreach \x in {2,...,4} \draw[rounded corners=2pt,->] (\x,0) -- (\x,\x) -- (\x-1,\x+1) -- (\x-1,7.5);
		\foreach \x in {-1,...,1} \draw (3+.2*\x,-1.4) node{$\cdot\mathstrut$};
		\foreach \x in {-1,...,1} \draw (2+.2*\x,9.4);
		\draw[rounded corners=2pt,->] (5,0) node[below]{$N$} -- (5,5) -- (4,6) -- (4,7.5);
	}  = \ordprod_{k=1}^{N-1} \mathbb{P}_{k \, k+1} 
\end{equation} 
and commutes with the hamiltonian \eqref{eq:susy_HS}: $[\mathbb{G}, H_{\text{s}\textsc{hs}}] = 0$. 

To describe further symmetries of this model it is prudent to reformulate the model using creation and annihilation operators, following \cite{Hal_94}: Let $a_i,a_i^\dagger$ be a set of canonical fermionic operators associated to site $i$, and $b_i,b_i^\dagger$ similarly a set of bosonic operators. These satisfy
\begin{equation}
	\label{eq:local_pre_fermions}
	\{ a_i, a_j^\dagger\} = \delta_{ij} \, , \quad [b_i, b_j^\dagger] = \delta_{ij}
\end{equation}
with all other brackets vanishing. Imposing the constraint that each site should contain precisely one particle (be it bosonic or fermionic) amounts to requiring
\begin{equation}
	\label{eq:constraint}
	a_i^\dagger a_i + b_i^\dagger b_i = 1\, , \quad \text{for all } i\, . 
\end{equation} 
The local two-dimensional Hilbert space on site $i$ is $V_i \coloneqq \C a_i^\dagger \ket{0} \oplus  \C b_i^\dagger \ket{0}$. Any non-trivial operator on $V_i$ must preserve the constraint \eqref{eq:constraint}, and so it makes sense to introduce the (fermionic) operators $f_i^\dagger \coloneqq a_i^\dagger b_i$ and $f_i\coloneqq  b_i^\dagger a_i$, which annihilate one excitation whilst creating the other. They satisfy
\begin{equation}
	\label{eq:local_fermions}
	\{ f_i, f_j^\dagger\} = \delta_{ij} \, , \quad \{f_i, f_j\} = \{ f_i^\dagger, f_j^\dagger\} = 0\, . 
\end{equation}
On the bosonic 'vacuum' $\ket{b} \coloneqq b_i^\dagger \ket{0}$ (with $f_i \ket{b} = 0$) these generate the same two-dimensional vector space $V_i = \C \ket{b} \oplus \C f_i^\dagger \ket{b}$. This is a  $\mathfrak{gl}(1|1)$-representation by the map $\mathfrak{gl}(1|1) \to \mathrm{End}(V_i)$ 
\begin{equation}
	N \mapsto f_i^\dagger f_i\, , \quad E \mapsto 1\, , \quad F \mapsto f_i\, , \quad F^\dagger \mapsto f_i^\dagger
\end{equation}
defined on the four generators of $\mathfrak{gl}(1|1)$, satisfying the non-trivial relations
\begin{equation}
	[N,F] = -F\, , \quad [N,F^\dagger] =F^\dagger\, , \quad 	\{F,F^\dagger\} = E
\end{equation}
with all other brackets vanishing. 

A way to realise the oscillators explicitly is by the Jordan-Wigner transformation, expressing the $f_j$ in terms of (bosonic) Pauli-matrices 
\begin{equation}
	f_j = \left(\prod_{k=1}^{j-1} \sigma^z_k \right) \sigma_j^+\, , \quad 	f_j^\dagger = \left(\prod_{k=1}^{j-1} \sigma^z_k \right) \sigma_j^-\, . 
\end{equation} 
One can check that identifying $\ket{b} \equiv \ket{\uparrow}$ and $f_i^\dagger \ket{b} \equiv \ket{\downarrow}$ indeed satisfy the expected local relations \eqref{eq:local_fermions}. 

The local spaces interact through the permutation operator $\mathbb{P}_{ij}$, which can be expressed in terms of $f_i$'s as 
\begin{equation}
	\mathbb{P}_{ij} = (f_i - f_j)(f_i^\dagger-f_j^\dagger)-1\, . 
\end{equation}

The whole system has a global $\mathfrak{gl}(1|1)$-symmetry: all four generators
\begin{equation}
	\begin{aligned}
	N_G &= \sum_{i=1}^N f_i^\dagger f_i\, , \quad E_G = 1 \, , \quad F_G= \sum_{i=1}^N f_i\, , \\
	F^\dagger_G &=  \sum_{i=1}^N f_i^\dagger
	\end{aligned}
\end{equation}
commute with $\mathbb{P}_{ij}$, and hence also with $H_{\text{sHS}}$. In terms of Pauli matrices $N_G = N/2-S^z$, and so the chain also has $S^z$-symmetry. 

The general $\mathfrak{gl}(n|m)$-Haldane-Shastry chain enjoys a further extended symmetry in the form of a representation of the Yangian $Y(\mathfrak{gl}(n|m))$ on the full Hilbert space, which can be expressed compactly in terms of oscillators (see \cite{Hal_94} eqn. (2.41)). For $n=m=1$ much of this extended symmetry becomes trivial: the would-be affine raising and lowering operators $Q^\pm$ act as zero on the Hilbert space, leaving only 
\begin{equation}
	Q^z = \frac{1}{2} \sum_{i< j} \cot \frac{\pi}{N} (i-j) \left( f_i^\dagger f_j - f_j^\dagger f_i \right)
\end{equation}
as an extra symmetry: $[Q^z,H_{\text{sHS}}] = 0$. Concretely, this means there are no extra spectrum-generating symmetries beyond $F_G$ and $F_G^\dagger$, and so the spectrum consists of $\mathfrak{gl}(1|1)$-doublets. Its highest weight states are characterised by their energy, momentum,
%$, and $Q^z$-eigenvalue
and the fact they are annihilated by $F_G$. Acting with $F_G^\dagger$ on this highest weight produces one descendant, any subsequent action annihilates due to $(F_G^\dagger)^2 =0$. Each highest-weight vector corresponds to a $\mathfrak{gl}(1|1)$-motif (or `supermotif'), a sequence of $l\leq N$ increasing integers $\nu_i \in \{1,\ldots,N-1\}$. Contrary to the Haldane-Shastry case (which is described by $\mathfrak{gl}(2|0)$ motifs as in \eqref{eq:motif_relation}) the $\nu_i$ may be only one apart. Given such a sequence, the associated doublet is labelled by eigenvalues
\begin{equation}
	%\label{eq:susy_eigenvalues}
	\begin{aligned}
	E_{\text{s}\textsc{hs}} &= \frac{1}{4}\sum_{i} \nu_i (N-\nu_i)\, , \quad P = \frac{2 \pi}{N} \sum_{i} \nu_i\, , \\
	 \lambda^z &= \sum_{i} (N/2-\nu_i)\, . 
	 \end{aligned}
\end{equation}
\vspace{0pt}
\bibliography{bibliography}

\end{document}